\UseRawInputEncoding
\documentclass[12pt]{book}
\usepackage[paperheight=234mm,paperwidth=150mm]{geometry}

\usepackage{xfrac}
\usepackage{graphicx}
\usepackage{mathptmx}
\usepackage{amsmath}
\usepackage{chapterbib}
\usepackage{pdfpages}
\usepackage{blindtext}
\usepackage{fancyhdr} 
\usepackage{parskip} 
\usepackage{gensymb}
\usepackage{textcomp}

\title{\Huge Aerodynamics\\
\bigbreak
\large at the\\
\huge Particle Level\\
\bigbreak
\large {\em A close look at some effects of particle flow\\ without the fluid approximation}}
\author{Charles A. Crummer, PhD}

\let\SavedIndent\indent
\protected\def\indent{%
  \begingroup
    \parindent=\the\parindent
    \SavedIndent
  \endgroup
}
\begin{document}
\frontmatter

\maketitle

\setcounter{chapter}{0}
\newpage
\begin{center}
Charles A. Crummer, PhD\\
\small {University of California, Santa Cruz (ret.)}\\ 
\scriptsize {charlie.crummer@gmail.com}\\
\date{\today}\\
Published through Amazon.com \\
Copyright \copyright 2018 by Charles A. Crummer \\
\vspace*{0,5in}
All rights reserved \\
\vspace*{2in}
Aerodynamics at the Particle Level / Charles A. Crummer.
\    Includes bibliographical references and index.\\
\    ISBN 978-0-692-16800-4
\    1. Aerodynamics--Subsonic. 2. Physical Sciences--Research. 3. Particle kinetics.

\vspace*{1in}
\textit {\small To Maryellen and my wonderful family \\ and \\ To Christine who helped me realize my dream}
\end{center}

\newpage
\chapter{Acknowledgements}
I would especially like to thank Professor George Brown at the University of California at Santa Cruz for encouraging me to study this subject and Professor Haraldur Sigurdsson at the University of Durham for his description of pyroclastic flow. Professor L. C. Woods` paper on the philosophy of science clarified certain important concepts. The under-appreciated work of Henri Coand\v{a} sparked interest in aerodynamic effects on surfaces. Professor Marco Colombini at the University of Genoa has done important work on mapping the pressure field around an airfoil. I also thank my friends Avis and Kendal Segr\` e for their friendship and encouragement.

\tableofcontents
\listoffigures

\chapter*{Preface}

{\it All} aerodynamic forces on a surface are caused by collisions of fluid particles with the surface.  Upwash, downwash, lift, drag, the starting vortex, the bow wave, and any other phenomena that would not occur without the surface are caused by its presence as it interacts with the air flow.  While the standard approach to fluid dynamics, which is founded on the ``fluid approximation,`` is effective in providing a means of calculating a wide range of fluid behavior, it falters in its ability to account for the effects of complex interactions of the fluid either with itself, other fluids, or with solid bodies.  One of the conditions required to justify the fluid approximation is that the flow be {\it steady} \cite{landau}, i.e. that the particles of the fluid not be interacting with each other or with any surface.  It is these very interactions, however, that are the causes of aerodynamic effects on solid bodies in the flow.  This is not to say, of course, that the fluid approximation is never useful, but that some well-known and important effects such as the Coand\v{a} effect are not explained by that model.

\mainmatter

\chapter*{Introduction}
The purpose of this paper is to set the stage for a close examination of fluid phenomena, an examination at the particle level.  Most fluid phenomena of interest are the result of its behavior in interaction with surfaces, other fluids or, indeed, with itself.  The eddies and turbulence attendant fluid shear are extremely complex.  As one fluid is injected into another, the shear effects depend further on the different attributes of the fluids.  If a fluid is flowing, it is doing so with respect to something, a surface for instance. 

A dimensionless quantity used to characterize the nature of fluid flow is Reynolds` number:

$$R = {\frac {\rho vL}  {\eta}}$$

where 

\begin{itemize}
\renewcommand{\labelitemi}{$\ $}
\item{$\rho$ is the density of the fluid,}
\item{ $v$ is its velocity,}
\item{ $\eta$ is the fluid\textquotesingle s viscosity and}
\item{ $L$ is called ``a characteristic length.``}
\end{itemize}

What does ``characteristic length`` mean?  $L$ is a length that is defined only in terms of the boundaries of the flow such as the diameter of a tube or the chord length of an airfoil.  What length is it and why?  In fact, Reynolds\textquotesingle number is only well-defined in discussions of model scaling of fluid flows in interaction with solid surfaces.  For example the characteristics of a flow around a boat with a beam of 4 meters in an ocean current of 10 knots will be the same for a scale model of the boat in the same ocean water whose beam is 0.4 meters and where the current is 100 knots. \\

What meaning can references to Reynolds\textquotesingle number have? \\

Bernoulli\textquotesingle s relation involves the fluid velocity.  In a Venturi tube, it is the velocity with respect to the wall of the tube.  If a high fluid velocity implies a low pressure, how can the pressure readings in different parts of the tube be different since the sensors are in the boundary layer of the fluid at the surface of the wall of the tube?  The boundary layer is stationary, or nearly so (see Figure \ref{fig:BL} below). \\

It is these and other baffling questions that have launched the author into these investigations.

Even though aerodynamics engineers are masters at designing airframes, they are refining known technology.  Without understanding from first principles, lighting engineers would just be refining incandescent lamps and we would not have fluorescent lights or LEDs.

\chapter{Background and Motivation}
The behavior of real fluids, i.e., compressible and viscous, is to this day baffling in many ways.  Part of the reason is that explanations of fluid behavior are hold-overs from the pre-twentieth century belief that a fluid is a fundamental entity, not composed of anything else.\cite{woods}  The trouble with this approach is that it provides only viscosity and pressure as ways of understanding how the fluid interacts with itself or with solid bodies.  Both are intensive variables but what do they mean in cases where the fluid approximation is not valid? \\

 Pressure, $p,$ (stress normal to a surface) can be understood as that fluid property which causes a normal force on a surface in the flow,

$$ d{\mathbf  F}_n =  p(s)\ d{\mathbf  A}. $$

The shear force provides part of the drag on a surface.  It is derived from the shear stress, ${\mathbf  \tau}$, tangential to the surface.  

$$ d{\mathbf  F}_s =  {\mathbf  \tau}\  \otimes \ d{\mathbf  A}, $$

where

\begin{equation}
\label{eq:visc}
{\mathbf  \tau} (s) \equiv \mu {{S(r,s)}\vline}_{r = 0}.
\end{equation}

Here, 

\begin{itemize}
\renewcommand{\labelitemi}{$\ $}
\item{$s$ is the location on the surface of the airfoil,}
\item{$r$ is a length in the direction normal to the surface, }
\item{${\mathbf  v}(s,r)$ is the velocity of the flow relative to the surface,}
\item{$\mu$ is the dynamic viscosity of the fluid, }
\item{$S(r,s) = {\partial{\mathbf  v}(r,s) / \partial{r}}$ is the shear and}
\item{$\tau$ is the resulting shear stress on the surface.}
\end{itemize}

\begin{figure}[!htb] 
  \centering
      \includegraphics[width=70mm]{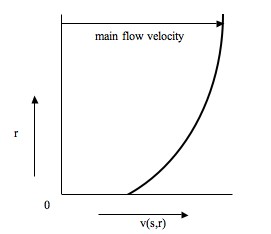}
      \caption{Velocity profile in the boundary layer for laminar flow}
      \label{fig:BL}
\end{figure}

 Figure \ref{fig:BL} shows qualitatively the velocity profile in the boundary layer during laminar flow.  The curve is differentiable and indicates that there is slip at the surface.  Admitting the possibility of slip at the airfoil surface is contrary to the no-slip assumption of Ludwig Prandtl\cite{schlichting} but in view of the development in Section \ref{sec:Mech} below and the work of Johan Hoffmman and Claes Johnson,\cite{Johnson} there is reason to suspect the reality of the no-slip assumption.  At the surface, because of the interaction of the particles in the flow with each other and with the (possibly submicroscopic) features of the surface, the behavior is very complex but for laminar flow this structure is smoothed out as the disturbance recedes into the flow.  \\
 
 A common example of this is the bow-wave of a slowly moving boat.   Close inspection of the behavior of the water at the bow reveals great complexity but far from the boat the wave is very regular and smooth.

In order that the viscosity, $\tau (s),$ as defined in Equation \ref{eq:visc} have meaning, the function ${\mathbf  v}(s,r)$ must be smooth and differentiable.  However, as the flow velocity increases, there is an onset of turbulence.  The boundary layer develops eddies near the surface~\cite{schlichting} and ${\mathbf   v}(s,r)$ becomes non-differentiable and so the partial derivative in Equation \ref{eq:visc} ceases to exist.  The behavior of the fluid becomes very complex and the flow becomes unsteady; the fluid approximation becomes invalid. \\

Since the work of Boltzmann~\cite{boltz} and Einstein~\cite{eins}, i.e., theory based on the postulate, and supporting evidence that fluids are composed of tiny particles, deeper insight is possible by considering in detail the interactions of these particles with each other, those of other fluids, and those of solid bodies in the flow.  In fact it may be helpful to remember that the only interactions a fluid can have, according to this model, are through momentum transfer or Van der Waals forces between its particles and between the particles and the surface.\footnote{We do not consider plasmas, which are affected by long-range electromagnetic forces.}  The molecules of a gas at standard pressure are only within Van der Waals distance about 1/100th of the time they are apart so these forces only play a part in particle-particle scattering. \\

The notion, therefore, that a streamline in a gas flow is ``attracted`` by a surface is not correct.  If a stream of gas, as in Coand\v{a} flow,\cite{coanda} seems attracted to a solid object it is due to its self-interaction, interaction with gas outside the flow, and the forces its particles exert on the surface as they strike it, not due to an attractive force between the particles and the surface.  Evidently, then, the ``attraction`` is due to the fact that a low pressure is created in the boundary layer and it is the higher ambient pressure that forces the boundary layer to stay in contact with the surface. In contrast to the work of Bernoulli, there is no ``Coand\v{a} equation`` because, other than Newton\textquotesingle s laws, we have no physical model for the behavior of the particles in the boundary layer.   Henri Marie Coand\v{a} was an engineer and observed effects that are widely incorporated into modern aerodynamic design but physicists have not developed a tractable mathematics to describe the behavior of such a large number, $\sim10^{23}$, of simple interactions without the fluid approximation.  In any case, to explore a mathematical model is not the same as to explore the physical world (See Appendix \ref{woodsapp}).  One goal of theoretical physics is the calculation of the results of experiments, another is to understand why the world works as it does.  The miracle is that mathematics is as useful as it is in describing and predicting physical effects. \\

The statements made below about fluid flow are conclusions and hypotheses coming from a consideration of particles obeying Newton\textquotesingle s laws.  The author\textquotesingle s intention is to stimulate the reader\textquotesingle s thoughts about the behavior of fluids in regimes where the flow is not steady, and hence the fluid approximation is invalid.  Another aim of this paper is to discern {\em causes} of phenomena.   A mathematical equation does not contain causal information.   For example, the thrust of a rocket is not {\em caused} by the velocity of the exiting gases but by the pressure difference between the throat and the projection along the axis of the motor of the throat area onto the back wall of the motor.  Bernoulli\textquotesingle s equation relates the the exit velocity and the pressure difference but conveys no information as to which is the cause and which is the effect.  It is only from experience with the physical world and abstractions of that experience that one knows that in Newton\textquotesingle s second law it is force that causes acceleration, not the reverse.  \\

It is hoped that an understanding of the true causes of aerodynamic effects will lead to new aerodynamic designs and the rethinking of designs already created.  Imagine, for a moment, that in the absence of a tractable mathematical model, non-mathematical understanding is possible. 

\chapter{Total force on the surface of the airfoil}
For perfectly elastic collisions the effect on a surface over an area $\Delta {\mathbf  A}$ results in a force, $\Delta {\mathbf  F}$ with components normal and transverse to the area.   

$$ \Delta {\mathbf  F}  = m \sum_{\Delta A} {\mathbf  a}_i$$

where $m$ is the mass of one particle and the ${\mathbf  a}_i$ are the accelerations of the particles hitting the surface area $\Delta {\mathbf A}$ and the summation is over the area.  The normal components of the ${\mathbf a}_i$\textquotesingle s are due to pressure and the transverse components are due to the viscous interaction of the fluid with the surface and with itself. \\

As the particles move over the surface, they are affected by the molecular protuberances on the surface and by Van der Waals forces between the particles and the surface.  This friction, i.e., viscosity, force is proportional to the area $\Delta {\mathbf  A}$ as well.

The total force on the airfoil, then, is the vector sum of the normal and tangential force components over the total airfoil area:

$${\mathbf F}_{total} =  -\sum_{airfoil} \left({ \Delta {\mathbf F}_n +  \Delta {\mathbf F}_s}\right).$$

The integral form of this equation is

\begin{equation}
\label{eq:Ftotal}
{\mathbf F}_{total} =  -\oint_{airfoil} \left({d {\mathbf F}_n +  d {\mathbf F}_s}\right).
\end{equation}

The minus sign indicates that the force on the particles is opposite to the force on the surface of the airfoil.  There is sometimes confusion about the above surface integral.  It does not mean that there is a net forward flow under the airfoil. Just that the flow is the vector sum of a translation and a rotation. This rotation is a consequence of the presence of the airfoil in the flow. The surface integral, reduced to a line integral in the case of an infinite wing, just indicates the integral of all forces on all sides of the wing.  (See also Section \ref{lift_calc} below.)  There is a formal derivation of the Kutta-Joukowski theorem at \cite{KJtheorem} and \cite{KJderivation}. Notice that the circulation integral is conducted in a region of potential flow, where Bernoulli\textquotesingle s equation would hold, i.e., far from the surface.  The circulation is an integral of the velocity field around the airfoil.  But the Kutta-Joukowski theorem begs the question as to how this velocity field is created and how it could be derived mathematically from first principles.

\section{Physical parameters affecting the pressure on the airfoil}

We will ignore Van der Waals forces and assume that the fluid is approximately an ideal gas. The equation of state for an ideal gas is

\begin{equation}
\label{eq:IDgas}
pV  = NkT,
\end{equation}

where

\begin{itemize}
\renewcommand{\labelitemi}{$\ $}
\item{$p$ is the pressure,}
\item{$V$ is volume,}
\item{$T$ is the Kelvin temperature.}
\item{$N$ is the number of particles in $V$ and}
\item{$k$ is Boltzmann\textquotesingle s constant, ($\sim 1.38 \times 10^{-23} {\sfrac {(m^2 kg)} {(sec^2 \ ^\circ K)}}$})
\end{itemize}

We now choose units so that $k = 1.$ Then the particle density, $\rho$ is

\begin{equation}
\label{eq:density}
\rho = {\frac N  V} = {\frac p  T}
\end{equation}

Then, 

\begin{equation}
\label{eq:pressure}
p = \rho T.
\end{equation}

Far away from the airfoil, the pressure, $p,$ is approximately constant and uniform except for the effect of gravity and the presence of the airfoil\cite{KJderivation}.  The compressibility of air can be ignored.  But {\em on the surface} of the airfoil, it is precisely pressure {\em differential}  that causes lift.  Equation \ref{eq:pressure} reveals that $\rho, $ and $T$, subject to the laws of thermodynamics, are at the disposal of the aeronautical engineer for creating a favorable pressure field on the airfoil\textquotesingle s surface.

\chapter{Mechanics of fluid interaction}\label{sec:Mech}
Aerodynamic forces affecting a rigid surface are always net forces produced by differences in pressure between different parts of the surface.  The absolute pressure on a surface area element is the density of the normal components of the forces acting on the surface there.

\begin{figure}[!htb] 
  \centering
      \includegraphics[width=70mm]{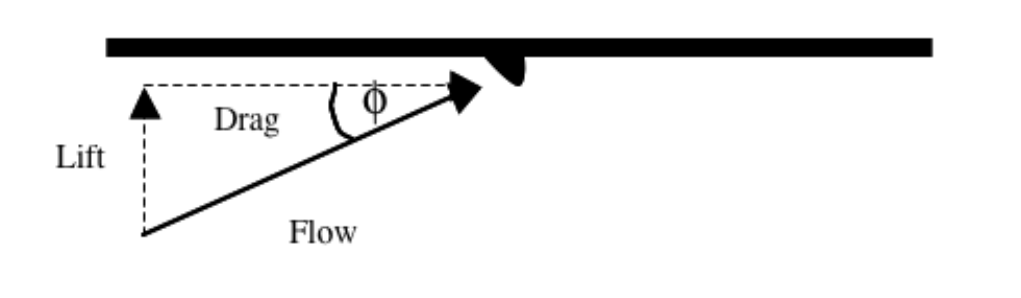}
      \caption{Interaction between fluid particles and	 a real surface}
      \label{fig:Drag}
\end{figure}

Aerodynamic forces on a body are caused {\it only} by collisions of fluid particles with the body\textquotesingle s surface.\footnote{The Coand\v{a} effect in liquid-surface flow, however, may be caused in large part by van der Waals forces, which are attractive.}  At the molecular level, the flow particles encounter any surface as a molecular structure which is rough, with protuberances whose size is of the order of magnitude of the flow particles themselves (see Figure \ref{fig:Drag}.).  As particles collide with the surface, their momentum components normal to the surface there cause lift, positive or negative, and stagnation pressure and the parallel components cause viscous drag and give rise to a boundary layer which is carried along by the surface (see Ref. \cite{schlichting}).  It is clear, then, that the microscopic structure of the surface and the properties of the fluid will affect drag and lift, even for $\Phi = 0$.\footnote{In the design of his racing plane, the H-1, Howard Hughes insisted that the rivets attaching the aluminum skin be flush rather than projecting above the skin.} A perfectly smooth surface would have no viscous drag, there would be no shear in the fluid near the surface and, it would appear, a wing made of this material would have lift only if the air flow momentum density had components normal to the bottom surface of the wing, i.e. due to the angle of attack, $\Phi$.  

Even though these momentum transfers occur only in the boundary layer that appears to be ``dragged along`` by the surface, they are responsible for the whole of lift and drag.  Actually, fluid particles can leave and enter the boundary layer by moving normal to the surface.  Dust on a surface in a flow is not disturbed laterally because the boundary layer is motionless, or nearly so, at the surface.  The boundary layer is created by the interaction of the main flow particles with particles bouncing off the surface.  For the time being, we assume that all collisions, particle-particle and particle-surface, are perfectly elastic and that the particles are spheres.

\subsection{Fluid flow over a flat surface}
\begin{figure}[!htb] 
\centering
      \includegraphics[width=70mm]{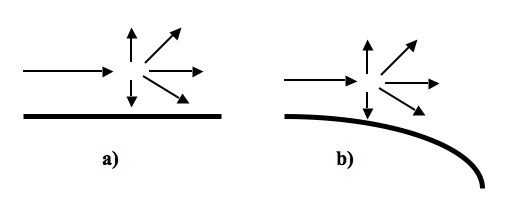}
      \caption{Fluid flow over curved and flat surfaces}
      \label{fig:CurvedSurface}
\end{figure}

Let us consider the flat surface in panel a) of Figure \ref{fig:CurvedSurface}.   The pressure on the surface is due only to the normal components of the momenta of the impacting particles.  Flow along such a surface will not affect surface pressure.  As particles are blown away from the surface, other particles are drawn in from outside to replace them.\footnote{Place a sheet of paper flat on your hands.  Blow over the top surface of the paper.  This experiment refutes the notion that the pressure in a free flow is less than the ambient static pressure.  Bernoulli flow, on the other hand, is (or could be) confined to a tube and is not free.}  Pressure on the surface is due to collisions of particles with the surface.  Where the flow has no normal component, the pressure is due only to the thermal motion and density of the particles in the boundary layer, i.e. the static atmospheric pressure.  Hence this pressure will be a function only of the mass of a particle, the particle density, and Kelvin temperature of the air at the surface.  
     
In reality, the fluid particles in a layer around a surface boundary seem to be carried along with the surface, i.e. the distribution of the components of their velocities parallel to the surface is nearly\cite{Johnson} circularly symmetric about a mean which is the velocity of the surface relative to the free-stream velocity.\footnote{This property of fluid flow was utilized by Nicola Tesla \cite{tesla} in his unique design of a rotary pump.}  Particles, as large as dust particles or as small as the molecules making up the flow, experience Van der Waals forces attracting them to the surface.  Whether or not the particles are fixed on the surface by these forces depends on the structure of the molecules making up the flow and those making up the surface.  These Van der Waals forces are responsible for the ``wetting`` of the surface.  In some cases, e.g. Teflon and water, the fluid drains off the surface quite readily just under the force of gravity.  In other cases, e.g. modern motor oil on a bearing surface, the fluid may adhere for months or even years.

In any case, however, particles continually leave the boundary layer and enter it transversely from the flow due to heat energy or, at an angle of attack, because they have velocity components normal to the surface.\footnote{It can be seen, then, that dust particles on a surface in an air flow are not disturbed not because the fluid particles are necessarily entrained but that they come and go normal to the surface.  Hence they do not impart lateral forces to the dust particles.}  Beyond a mean free path\footnote{$\sim9 \times 10^{-8}$ meters for $N_2$ at standard pressure and temperature.} or so but still in the boundary layer, the distribution of the normal components of particles` velocities moving toward or away from the surface will depend on the temperature and density of the particles.   Even though it is regularly driven at high speed, a car will accumulate dust on its body. An air stream directed toward the surface, however, will blow off some of that dust.  As we will see, it is the mutual interaction of flow particles and these ``stagnant`` boundary layer particles that is responsible for a part of the lift on an airfoil at subsonic speeds.

An increase in the free-stream velocity means that the components of the velocities of the flow particles increase in the direction of the free-stream velocity and parallel to the surface.  The reason that the boundary layer remains quiescent, or nearly so, is that the components of the colliding particles` velocities parallel to the surface reverse as they collide with microscopic irregularities.  This is one of the causes of aerodynamic drag and accounts for the fluid\textquotesingle s viscosity.\footnote{Though viscosity is supposed to be a property of the fluid, it is measured by the terminal velocity of a ball in the fluid or the force it takes to slide two plates with the fluid between them.  Viscosity, then, has to do with the interaction of the fluid with itself as well as with solid bodies.}  If the collisions are not perfectly elastic, the rebound speed is less than the incident speed and the surface absorbs some of the particle\textquotesingle s energy, i.e. it heats up. 

\subsection{Fluid flow over a curved surface}\label{Flfl}

In a steady flow over a surface, stream particles have only thermal velocity components normal to the surface.  If the surface is flat, the particles that collide with boundary layer particles are as likely to knock them out of the boundary layer as to knock others in, i.e. the boundary layer population is not changed and the pressure on the surface is the same as if there were no flow.  If, however, the surface curves away from the flow direction, the particles in the flow will tend to take directions tangent to the surface, i.e. away from the surface, obeying Newton\textquotesingle s first law.  As these particles flow away from the surface, their collisions with the boundary layer thermal particles tend to knock those particles away from the surface.  What this means is that if all impact parameters are equally likely, there are more ways a collision can result in a depletion of the boundary layer than an increase in the boundary layer population.  The boundary layer will tend to increase in thickness and to depopulate and, according to Equations \ref{eq:density} and \ref{eq:pressure}, the pressure will reduce there.  This is why the flow is forced toward the surface, the Coand\v{a} effect with the attendant ``suction`` that draws in fluid far from the surface.\footnote{This explanation suggests experiments exploring the structure at the edge of the main flow that is away from the wall.  The explanation of the mechanism by which the flow is ``attracted`` to the wall implies how the flow should behave at its other edge too.}  Those particles in the flow that do interact with the stagnant boundary layer will give some of their energy to particles there.  As they are deflected back into the flow by collisions with boundary layer particles, they are, in turn, struck by faster particles in the flow and struck at positive impact parameters.

\begin{figure}[!htb] 
\centering
      \includegraphics[width=70mm]{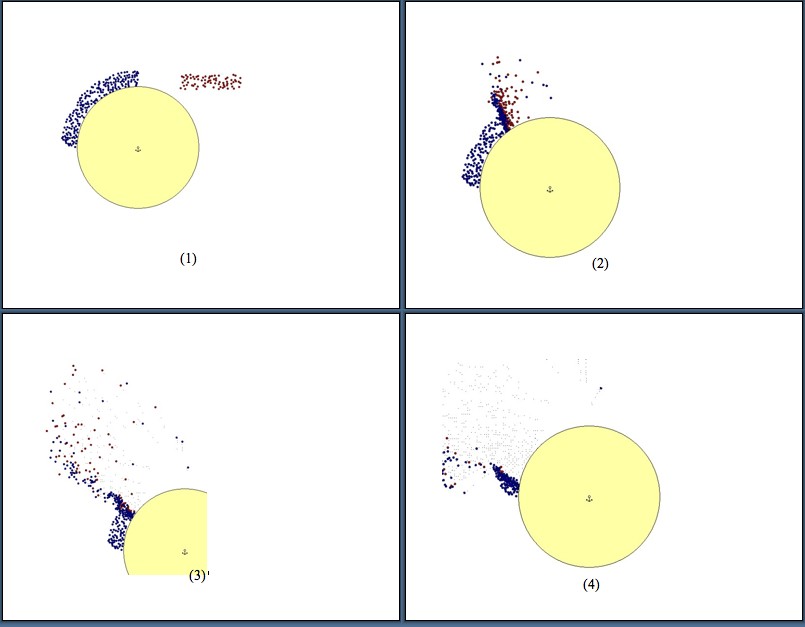}
      \caption{Behavior of particle flow over a curved surface}
      \label{fig:Coanda-flow}
\end{figure}

The following, Figure \ref{fig:Coanda-flow}, consists of 4 frames taken from an animation\cite{vianni} illustrating the behavior of flow particles as they interact with stagnant particles in the boundary layer.  The first panel shows the incoming particles in red approaching from the right.  In the second panel the incoming particles begin to interact with the stagnant particles meant to approximate a boundary layer. The third panel shows the boundary particles being blown away by the incoming set, thus reducing the pressure at the surface.  As panel four shows, it is primarily the boundary layer particles that make up the flow that clings to the curved surface.

These frames show, at least qualitatively, the Coand\v{a} effect.  The figures are frames taken from an animation made with Working Model\texttrademark\cite{WM} software.  In the video, approximately 600 small circles are launched toward a fixed circle with stagnant circles positioned around it, meant to simulate a boundary layer.  All collisions are perfectly elastic and the large circle has infinite mass.  Of course, this model is highly unrealistic because of the very small number of particles, their simple circular structure, the smoothness of the surface and the absence of thermal motion.  It does, however, show boundary layer depletion and the ``wrapping`` of the flow around the surface.  A more accurate simulation would have a continuous flow impinging on a rough surface surrounded by particles.  All the particles should be interacting thermally with each other and with the surface.  Notice also that the wrapped flow contains almost none of the red incoming particles.  

In reality, the flow shears past the surface (where the molecular motion is complex and chaotic) and the fluid velocity as a function of the distance normal to the surface is a smooth function of this distance.  The velocity profile curve parameters are constants depending on the fluid velocity, the physical characteristics of the surface and the particles making up the fluid (See Figure \ref{fig:BL}).  In any case, as the flow velocity increases, separation points will begin to appear\cite{schlichting}.  These are the points on the surface where the directional derivative of the fluid velocity normal to the surface vanishes as does the shear (Equation \ref{eq:visc}).  As the fluid velocity increases even further, the derivatives at the separation points actually reverse sign, there is backward flow on the surface.\cite{schlichting}  Vortices have formed downstream from these stagnation points.  All this is in the language of fluids.  What is going on at the particle level though? 
 
The curved part of the surface acts as the obstruction mentioned in the explanation of the vortex process (See Section \ref{sec:vortex}.) because it presents stagnant particles to the flow.  The flowing particles as they approach the surface interact with these particles and with the surface itself.  Some populate the boundary layer and then interact as stagnant particles with other particles in the flow.  There is a constant interchange of particles between the boundary layer and the flow.  As these interactions take place the process described above activates the boundary layer particles like falling dominoes, causing the enveloping flow.  When the surface curves away from the flow, the flow particles, obeying Newton\textquotesingle s first law, tend to travel on trajectories tangent to the surface and thus leave its vicinity, taking some boundary layer particles with them.  This reduced pressure in the boundary layer has two effects.  First, it causes the higher pressure in the main flow to force itself, and smoke streamers, toward the surface, and second, it results in lift as the higher pressure on the bottom of the wing has increased effect.   

\begin{figure}[!htb]
\centering
\includegraphics[width=70mm]{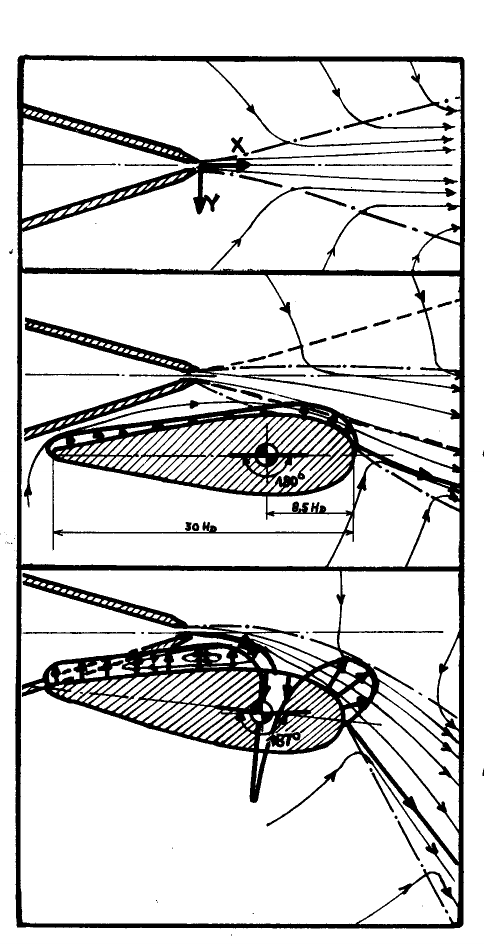}
\caption{Flows over a Wing}
\label{fig:WingFlow}
\end{figure}

The Coand\v{a} effect is investigated in some detail in articles in {\it Deutsche Luft- und Raumfahrt}\cite{deutsche}.   H. Riedel\textquotesingle s paper has numerous diagrams of flow patterns and distributions of pressure differentials on a wing surface in various positions with respect to an air jet.  Figure \ref{fig:WingFlow} is from this paper.  

The first panel in Figure \ref{fig:WingFlow} shows the flow of and around a free stream in an atmosphere.  Note the entrainment of air from outside the stream.  Panels 2 and 3 show flow and distributions of pressure differentials on a wing in the flow.  Note that in panel 3 there is a sharp spike in downward pressure where the flow actually impinges vertically on the surface.  This downward pressure is the cause of wing stall.  Flow  along a {\em positively curved} surface causes a lowering of the pressure on the wing but for flow rates above a certain velocity a vortex will be created which can cause an increase of pressure there. 

The Coand\v{a} effect gives a hint at at what turns out to be the most important cause of lift at zero or small angle of attack and subsonic conditions.  While it may be intuitive that to increase lift the pressure under the wing, and thus the angle of attack, should be increased, in fact for a fixed wing it is more important to  {\em decrease} the pressure on the {\em top} of the wing. (See Reference \cite{anderson}, page 181.)  Professor Marco Colombini at the University of Genoa, Italy\cite{colombini} has produced some interesting animations illustrating the pressure distribution around a standard airfoil at varying angles of attack.\footnote{These pressure distributions, however, do not show the bow wave the same as it is seen in Figure \ref{fig:forces_hepperle}.}   It is interesting to think of airfoil design as an exercise in {\em managing buoyancy.}

\subsection{Static buoyant lift: the aerostat}\label{subsec:Aerostat}

Dirigibles, helium balloons and hot-air balloons utilize buoyant lift.  They are sometimes called aerostats because they achieve lift without movement, without a main air flow.  They generate neither upwash or downwash as a third-law reaction to this lift.  These devices rise due to the difference in the atmospheric  pressure between top and bottom.  This buoyant force acts naturally on everything immersed in a fluid in a gravitational field.  It acts on us but we don`t notice it because the density of our bodies is so much greater than the density of air where we live.  Archimedes noticed this buoyant force and uttered the famous ``$E \upsilon \rho \eta \kappa \alpha $\ !``  He had discovered how to measure the density of the king\textquotesingle s crown and to test if it was pure gold.

One might think that the buoyant force, which is due to the gravitational field, would be negligible for an airplane because the airplane\textquotesingle s overall density is much greater than air at standard conditions.  However, see Section  \ref{sec:coanda_effect} below.  Since the buoyant force is due to the pressure differential between the top of a body and the bottom as well as the overall density of the balloon, the buoyancy can be managed by controlling these pressures and the density.

In France a balloon large enough to carry cargo is called a Montgolfi\`ere after the brothers who in 1783 flew the first hot-air balloon to carry a living cargo.  

A balloon, or aerostat, consists of an envelope filled with a gas that is lighter than air, i.e. less dense.  This gas can consist of molecules each of which is lighter than the average weight of the molecules of air or air itself that is hotter and less dense than the atmosphere surrounding the aerostat.  The light molecules of Helium or Hydrogen are used for the former type of aerostat.

\subsubsection{Light gas Balloon}

The ideal gas equation, Equation \ref{eq:density} above, is:

\begin{equation*}
\rho = {\frac p  T } = {\frac N  V},
\end{equation*}

where $\rho$ is the number  density of the molecules, i.e., the number  of molecules per unit volume, and we`ve chosen units so that $k = 1.$

Archimedes` equation for the buoyant force, $B$,  on an object immersed in a fluid is 

\begin{equation}
\label{eq:Archi}
B = \rho_f m_f g V,
\end{equation}

where 
\begin{itemize}
\renewcommand{\labelitemi}{$\ $}
\item{$\rho_f$ is the density of the particles of fluid in which the body is immersed,}  
\item{$m$ is the average mass of a fluid particle,}  
\item{$V$ is the volume of the immersed object and}
\item{$g$ is the acceleration due to gravity.}
\end{itemize}

  This upward force is offset by the weight of the object immersed, 

\begin{equation}
w_{balloon} = \rho_g m_g 
g V + w_{envelope},
\end{equation}

where here $\rho_g$ is the density of the gas in the balloon, $m_g$ is the mass of a gas particle and the $V$s are the same.  The net upward force, then, is

\begin{equation}
F_{net} = B - w_{balloon} = (\rho_f m_f - \rho_g m_g )g V. - w_{envelope}.
\end{equation}

Assuming that the temperatures of the gas inside the balloon and the outside the balloon are equal, for a balloon with a loose, light, flexible and very voluminous envelope, the atmospheric pressure and the pressure inside the balloon are the same so, according to Equation \ref{eq:IDgas},
 
\begin{equation}
\rho_f = \rho_g = \rho,
\end{equation}

and the net upward force is proportional to $m_f - m_g,$ (ignoring the weight of the envelope), i.e.,

\begin{equation}
F_{net} \propto ( m_f - m_g).
\end{equation} 

This is why a light gas like Hydrogen or Helium is used in high-altitude balloons.  

If the envelope of the balloon is an elastic material like rubber, the pressure exerted by the envelope on the gas within is an increasing function of the volume of the balloon.  The pressure inside the balloon is $p_g = p_f + p_{envelope},$ which means that in this case $\rho_g > \rho_f.$ A rubber balloon will only rise if $\rho_g m_g < \rho_fm_f$, and then only until $\rho_f  = \rho_g ({m_g / m_f})$ or the elastic limit of the rubber is reached and the balloon explodes. 

\subsubsection{The Hot-air Balloon}

The latter form of aerostat mentioned above, a hot-air balloon, is lifted by an envelope containing heated air.

\begin{figure}[!htb]  
\centering
\includegraphics[width=80mm]{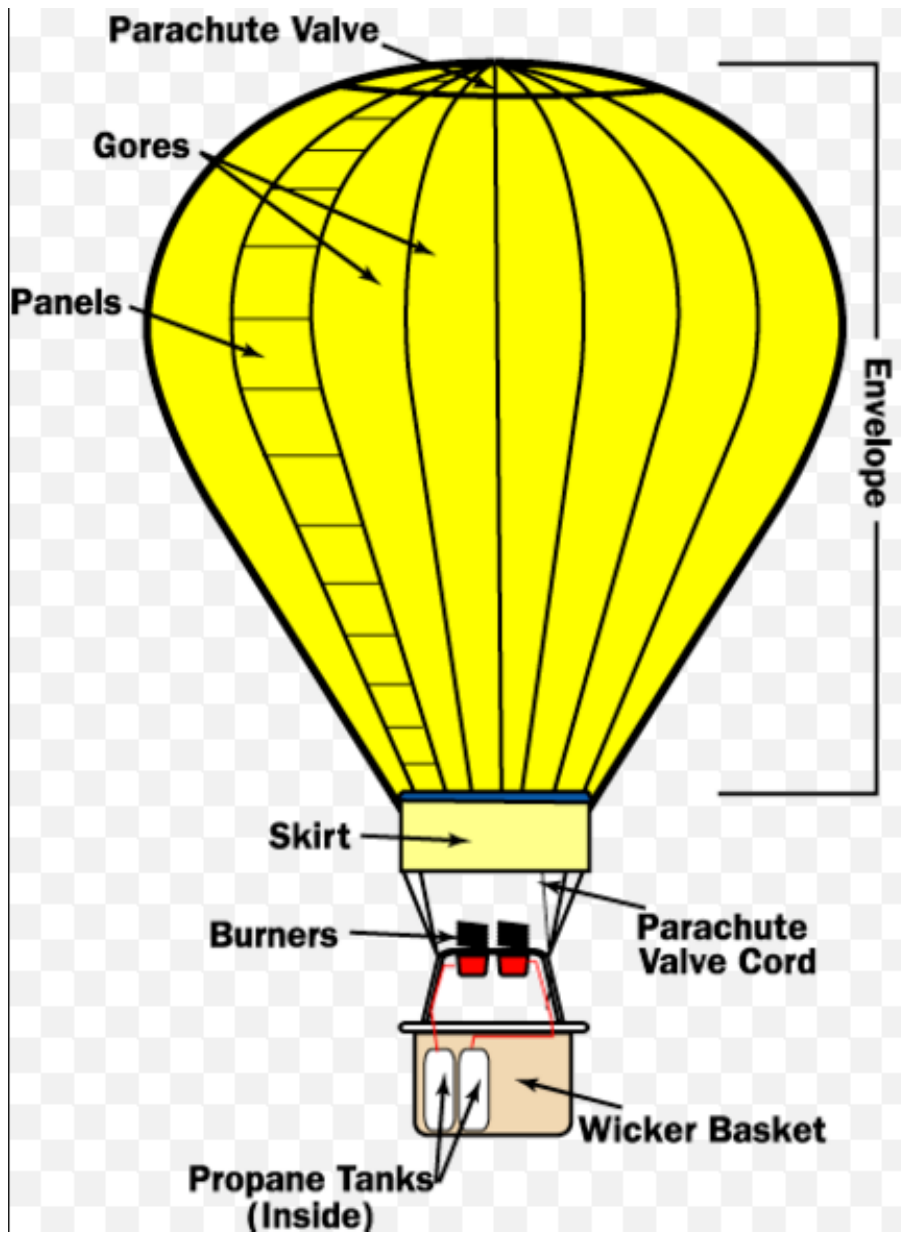}
\caption{Hot-air Balloon}
\label{fig:HaB}
\end{figure}

  The burner that provides the heat is concentric with the opening at the bottom of the balloon so that the hot air created by the burner rises and the cooler air that is displaced exits from the annular space around it and, if it is opened, the parachute valve at the top of the balloon. As the temperature increases, the pressure required to keep the envelope inflated is provided by fewer and fewer molecules and thus the weight of the balloon decreases. By the intermittent operation of the burner and the parachute valve, the pilot can very precisely choose the height of the balloon.

All the forces that result in the lift force, buoyant or dynamical, are due to collisions of molecules with the envelope.

\subsubsection{Submarines and Fish}
Both fish and submarines must be able to control their densities so they can adjust their depths in water.  Fish do this by changing $p$ in Equation \ref{eq:density}.  They do this by contracting or relaxing the muscles around their swim-bladders.

\begin{figure}[!htb]  
\centering
\includegraphics[width=80mm]{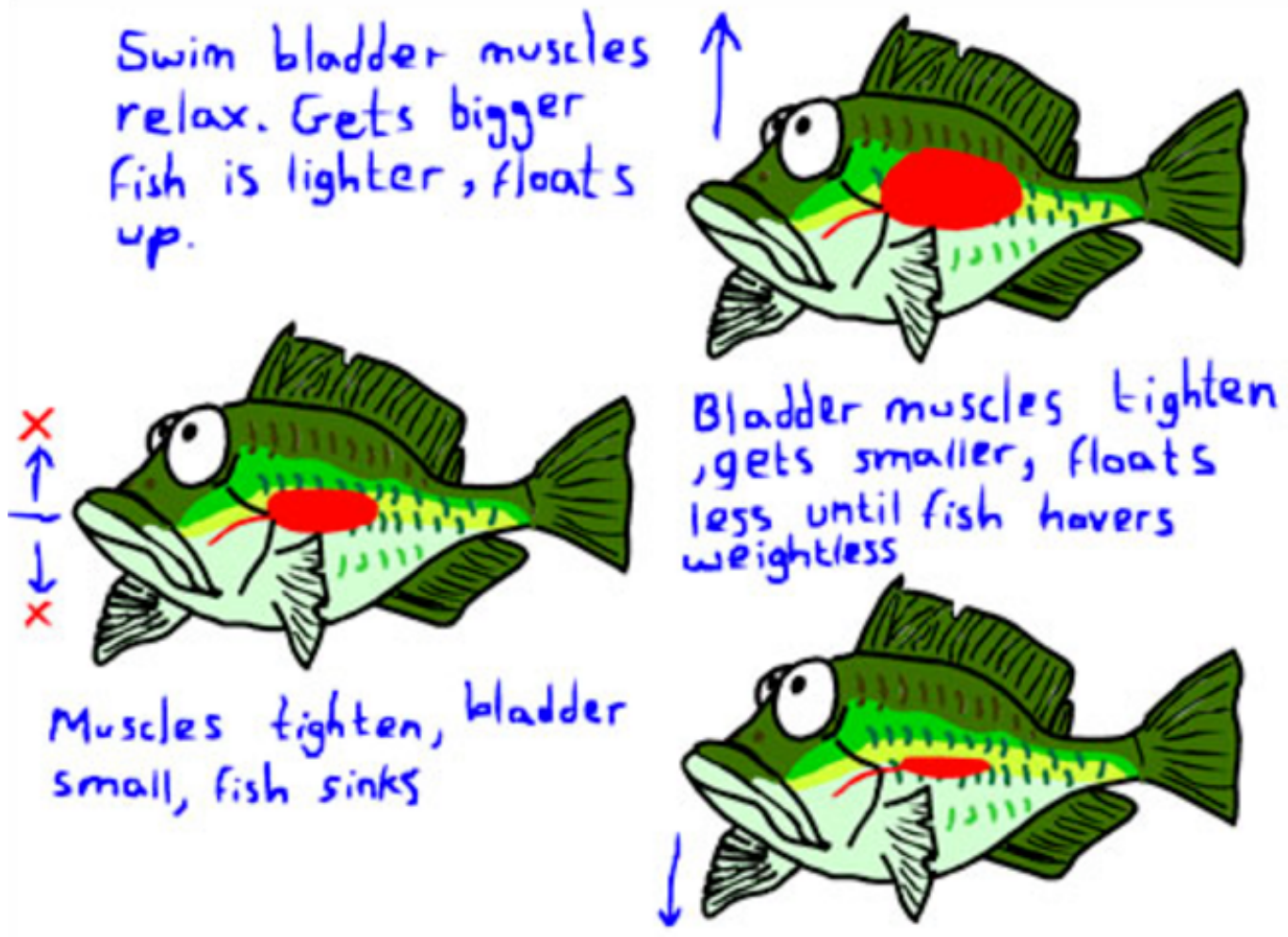}
\caption{Fish\textquotesingle s Swim-Bladder}
\label{fig:Swim-Bladder}
\end{figure}

A submarine\textquotesingle s density is changed basically by adjustments of the total weight of the craft.  It also has trim tanks, tanks that are the counterparts to ``swim-bladders,`` and ballast tanks for gross depth control.

The trim tanks operate much like a fish\textquotesingle s swim-bladders.  There is a more or less fixed quantity of air in them.  Their weight as well as the pressure of the air in them is adjusted by  injecting water or pumping it out.  

To cause the submarine to  submerge, the captain orders the ballast tanks, initially filled with air, to be partially flooded with water as air is exhausted. In this way, $N$ is reduced and the weight is increased as water floods the ballast tanks.  To surface again, the captain orders that  air from high-pressure  tanks be let into the ballast tanks to force the water out.  The weight of the  craft is increased and decreased by the taking-on and exhausting of seawater. This control is achieved both by the release of the high-pressure air into the ballast tanks and by pumps.  The pumps control both the ballast pressure and the overall weight of the vessel. 

In order for the pumps to work, a certain amount of air must remain in the ballast tanks at all times.  By the same token, the high-pressure tanks must be maintained at a pressure sufficient to overcome the water pressure at the deepest level the submarine is allowed to go.

\begin{figure}[!htb]  
\centering
\includegraphics[width=80mm]{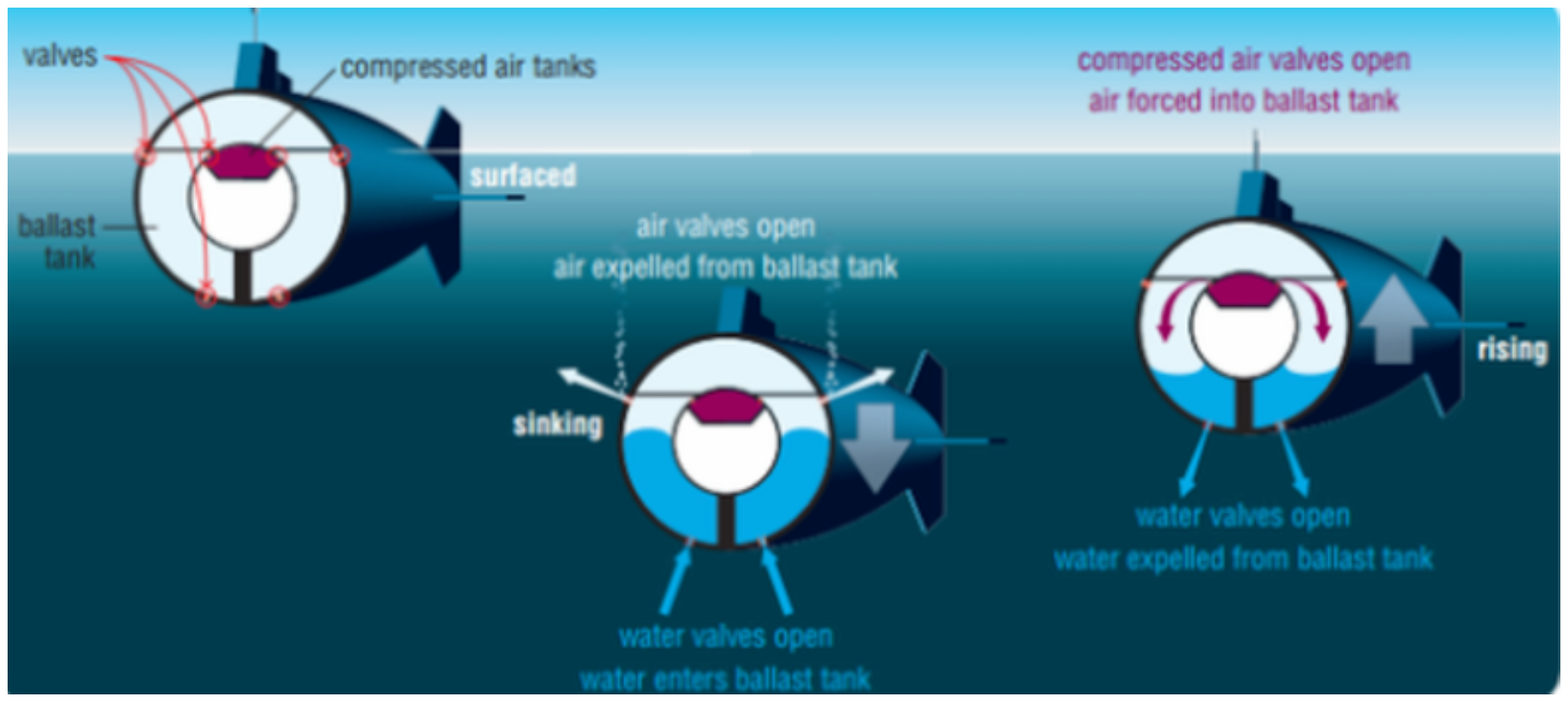}
\caption{Submarine Ballast System}
\label{fig:Sub-Ballast}
\end{figure}
\newpage

\begin{center} *********** \end{center}

In addition to its own weight a container of gas has weight due to the gravitational force which affects each molecule of the gas inside. In the gas there is a density gradient toward the earth. Newton\textquotesingle s law of gravitation is:

\begin{equation}
\label{eq:Newtongrav}
{\mathbf F} = -{\mathbf r}Mm/r^3,
\end{equation}             

where ${\mathbf r}$ is the vector from the centre of the earth to the particle.
   
 As they collide with one another molecules, in general, will have momentum components in any given direction due their temperature but they will always have a downward momentum component due to their acceleration in the gravitational field.  

\begin{equation}
m{\mathbf v} = m({\mathbf v}_{heat} + {\mathbf v}_{down}).
\end{equation}

This downward component will always be a factor in collisions with other particles and will be transmitted in each collision according to Newton\textquotesingle s third law. The momentum transfer to the envelope from the molecules on the bottom will be greater than the molecules at the top just because of the greater particle density there. The top molecules exert a force upward and the molecules on the bottom exert a downward force. In a vacuum the difference between these forces is that part of the weight of the object that is due to the gas inside. 

If the container is in the earth\textquotesingle s atmosphere there is also a buoyant force. This force is also due to the gravity force. The external atmosphere is denser at the bottom of the envelope than at the top, hence the number of exterior atmospheric molecular collisions per second at the bottom of the envelope is greater than at the top. The difference between the number and force of the collisions per second on the bottom (molecules with upward velocity components) and that on the top (downward components) is the buoyant force. If the buoyant force is greater than its total weight, the container rises. 

The number density, $\rho (y),$ of molecules inside the envelope is a monotonically decreasing function of  the height. This is due to the influence of gravity on each molecule. If hot air molecules are injected into the envelope those molecules will undergo collisions with the molecules they encounter.  

Consider a hot molecule that enters with energy,

\begin{equation}
\Delta E = k \Delta T,
\end{equation}  

where $k$ is Boltzmann\textquotesingle s constant and $\Delta T$ is the temperature of the entering molecule.  Again, we choose units for which $k = 1$. 

Consider two regions: upper and lower.  In each, according to Equation \ref{eq:density},

\begin{equation}
p = \rho T.
\end{equation}

Although the densities and temperatures of the two regions differ, there is only one pressure, $p$, at the point of entry.  Hence,

\begin{equation}
\rho_u \times T_u = \rho_d \times T_d,
\end{equation}  

and

\begin{equation}
T_u = {\frac {\rho_d}  {\rho_u}} \times T_d.
\end{equation}  

This molecule\textquotesingle s energy will be shared among the molecules it collides with and their added energy will be shared as they interact with others.

Assume the molecule is knocked upward. The resulting increase in temperature in that region, $u$ , compared with the lower region, $d$, will be,

\begin{equation}
\Delta T_u = {\frac {\rho_d (y)}  {\rho_u (y)}} \times \Delta T_d,
\end{equation}  
 
and similarly for the case where the entering molecule is knocked down.  If the entering molecule is knocked upward, there will be fewer molecules to share its energy than if it were knocked downward since, due to the gravity field,

\begin{equation}
\rho_u (y) < \rho_d (y).
\end{equation}     

The absorption of its energy there will cause a rise in temperature greater than if it were knocked downward.  If it is knocked downward, the denser aggregation of molecules there will also absorb the molecule\textquotesingle s energy but the rise in temperature will be less. It is not just that the hot air molecules \textit{themselves} rise, it is also their \textit{energy} which is transmitted upward as they collide with molecules already there.

\subsection{Vortex fluid motion}\label{sec:vortex}
As a fluid stream passes through an opening in a barrier into stagnant fluid, eddies appear.  Consider the state of the fluid as the flow begins.  Behind the barrier the distribution of velocities of the particles of the fluid in a small volume is spherically symmetric (except for the effect of gravity) and the mean of the distribution is a function of the Kelvin temperature.  

\begin{figure}[!htb] 
\centering
      \includegraphics[width=70mm]{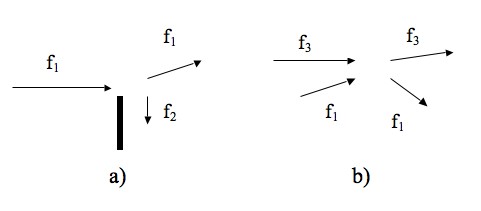} 
      \caption{The vortex process}
      \label{fig:VortexProcess}
\end{figure}

Upstream, the pressure behind the barrier is higher than the pressure behind the exit.  A particle on a streamline just grazing the barrier encounters particles behind that barrier whose mean velocities are zero.  Downstream of the barrier, the result of collisions with these stagnant particles is the slowing of a flow particle as well as its deflection back into the flow.  (See Figure \ref{fig:VortexProcess}.)   

\begin{figure}[!htb] 
\centering
      \includegraphics[width=80mm] {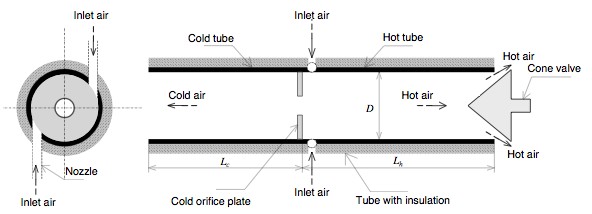} 
      \caption{Vortex tube schematic}
      \label{fig:VortexTubesch}
\end{figure}

\begin{figure}[!htb] 
\centering
      \includegraphics {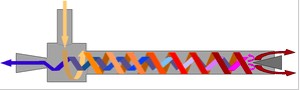} 
      \caption{Vortex tube flow}
      \label{fig:VortexTubediag}
\end{figure}

The greater the difference between the flow velocity and the thermal velocities of the stagnant particles, the closer to 90$^\circ$ from the flow direction will be the directions of the stagnant particles after the collisions.   Thus the interaction between the stream and the stagnant region serves to sort out the colder stagnant particles and force them away from the flow.  The vortex heat pump described later in Section \ref{VortexRefrig}, Figures \ref{fig:VortexTubesch} and \ref{fig:VortexTubediag} uses this principle. 

As the stream particles that have suffered collisions with stagnant particles are hit by faster ones in the stream, they too are deflected with a velocity component normal to the stream velocity.  As they continue after being deflected away from the stream, they hit other stagnant particles (Figure \ref{fig:VortexProcess}), forcing them toward the same center.  The result is that part of the flow is changed into a vortex.  If the obstruction is a hole in a plate, some of the energy of the flow is trapped in the form of a vortex ring.  If the flow is a pulse, this ring follows in its wake.

Let the lower half of the {\it y-z} plane be a barrier in the fluid.  (See Figure \ref{fig:VortexProcess}.)  The velocity of the flow will be superimposed on the random motion of the molecules, i.e. heat.  As the flow begins, say from minus to plus in the $x$-direction, the mean of the distribution of the velocities of those particles in the flow will be shifted toward positive $v_{\it x}$.  As these particles pass the barrier, they collide with fluid particles that have a velocity distribution with zero mean, i.e. the stagnant particles.  

Call an impact parameter positive if the location of the impact point with a particle in the flow is a positive distance in $y$ from the center of one of these particles.  If it passes the edge of the barrier with sufficient speed, a flow particle is likely to hit a particle behind the barrier with a positive impact parameter.  This will result in the stagnant particle being knocked back behind the barrier (See Figure \ref{fig:VortexProcess}a.).  The flow particle will be deflected up into the flow, with reduced momentum where it will be deflected by other particles in the flow and eventually be knocked back, away from the flow (Figure \ref{fig:VortexProcess}b).  These particles still have an {\it x}-component of velocity that is larger than their {\it y-z} velocities but as they interact with each other and other particles in the flow in the way described above, they will participate a circular flow and their energy will decrease.  This process repeated statistically with various impact parameters results in a vortex.  We will call it the {\it vortex process.}

Consider a velocity coordinate system local to a flow particle that passes very close to the barrier and with its {\it x}-axis in the flow direction.  Initially, that system will be aligned with the coordinates mentioned above.  As time goes on after the particle has passed the {\it y-z} plane, the local system will, on average, rotate around its {\it y}-axis.  One can see in this way that the effects of this interaction with the stagnant molecules will propagate into the flow on the downstream side of the barrier.  The result is a vortex.

\subsection{Finite wings and wingtip vortices}
\begin{figure} 
  \centering
     \includegraphics[width=70mm]{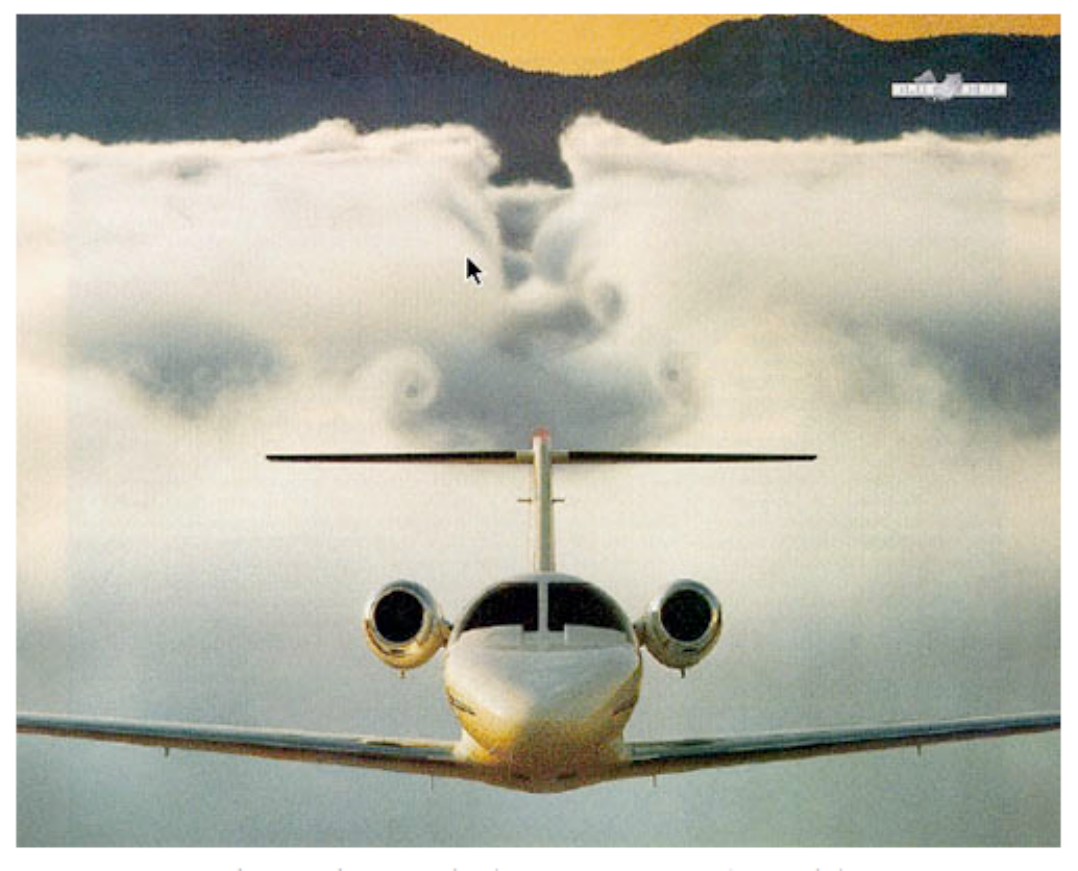}
     \caption{Downwash and wingtip vortices}
     \label{fig:wingtip_vortices}
\end{figure}

In flight, an airplane will generate a vortex at each wingtip.  These vortices are created as the higher pressure air under the wing leaks out from under the wing and away from the fuselage.  The vortex is formed as this air is drawn into the low pressure region above the wing.  The vortex is a nuisance\footnote{See http://en.wikipedia.org/wiki/Wingtip\_device\#NASA\_development for a description of devices to control the wingtip vortex.} and is a source of drag and instability due to the vortex impinging on the top of the wing there.  It can actually reduce the overall lift.  Figure
 \ref{fig:wingtip_vortices} shows the vortices very clearly.  Notice that the axes of these vortices are parallel to the flight direction.  The wingtip vortices are an unwanted effect and are due to the necessity of finite-length wings.
 
 Though the wingtip vortices are beautiful and spectacular, the concomitant to the most important factor in producing lift is the huge trench left in the cloud by the downwash off the trailing edges of the wings.
 
The French jet engine manufacturing company, Price Induction\footnote{Price Induction, 2, Esplanade de l`Europe 64600 Anglet, FRANCE.  www.price-induction.com} sells small high bypass engines for small aircraft.  One of their innovations is a turbofan using composite, non-metallic blades.  At speed, the fan blades elongate and actually seal on the special bearing surface of the fan housing.  The reason for this is to eliminate vortices at the vane tips.  This reduces power requirements, increases the engine efficiency and increases thrust.

\subsection{Leading Edge Extensions}
Though wingtip vortices are unwanted, similar vortices are created on purpose by so-called Leading Edge Extension (LEX) surfaces.\cite{LEX}   A LEX is a flat surface extending a short distance from the fuselage and from near the cockpit aft to the leading edge of the wing.  At angle of attack vortices are created as the high-pressure air flows from below the LEX to the lower pressure above.  This causes the vortices, clockwise on the left side and counter-clockwise on the right.  These vortices extend back over the wings and interrupt  the stalling vortices that would otherwise form over the wing.  They blow away the particles that would cause high pressure on the tops of the wings, especially near the roots.  LEXs  allow the plane to operate at higher angles of attack than it otherwise could. 

\subsection{Birds in flight}\label{birds}
The high-speed camera shows some very interesting aspects of birds taking flight.\cite{birds}  Perhaps the most interesting is that on take-off, when maximum lift is needed, a bird\textquotesingle s power stroke is down and {\em forward}, not backward as it would do if it were ``swimming`` in the air.  This motion both pressurizes the air under the wing and creates upwash\footnote{See Section \ref{slots}} on the leading edges of its wings.  This upwash flows over the leading edge and actually contributes to lowering the pressure on the top of the wing.  

On aircraft, the leading edge slots and slats are designed to control and make use of upwash.  Trailing edge flaps act like the big feathers on the trailing edges of a bird\textquotesingle s wings.  They help trap the flow and thus increase pressure under the wing and they also extend the wing\textquotesingle s curved surface and hence the region of low pressure on the top of the wing.    

\section{Bernoulli flow and Coand\v{a} flow}

\subsection{Bernoulli\textquotesingle s equation}
For an incompressible\footnote{See, however http://www.efunda.com/formulae/fluids/bernoulli.cfm for a more general form of the equation which describes the behavior of certain types of compressible fluids.}  fluid in {\it steady}\cite{landau} flow, a simple expression for the conservation of energy was derived by Daniel Bernoulli in 1737 in his book {\it  ``Hydrodynamica``}.  In steady flow, the fluid can be enveloped in an actual or virtual tube.  That means that at any cross-section perpendicular to the tube\textquotesingle s walls, the fluid has a uniform velocity across the tube, i.e. there can be no shear in the fluid.  Fluid neither leaves nor enters through the wall of the tube and the particles do not interact with each other or with the wall of the tube.  And finally, since the flow must be laminar, the tubes themselves, actual or abstract, are restricted to a smooth, gently varying shape.  These assumptions preclude turbulence or eddy formation.  If these conditions hold to a good approximation, Bernoulli\textquotesingle s equation holds.  If such a tube cannot be drawn, the equation does not hold.  Bernoulli\textquotesingle s equation allows the calculation of general behavior but because of these assumptions the theory is not able to predict other aspects of fluid dynamics, such as behavior in the boundary layer of a surface in the flow. 
 
Bernoulli\textquotesingle s equation is:

\begin{equation}
\label{eq:enden}
Energy~Density= {\frac 1  2} \rho{v^2} + \rho{gh} + p.
\end{equation}

where

\begin{itemize}
\renewcommand{\labelitemi}{$\ $}
\item{$p$ is the absolute pressure,}
\item{$\rho$ is the density of the fluid,}
\item{$g$ is the acceleration due to gravity,}
\item{$h$ is the height in the gravitational field and}
\item{${\mathbf v}$ is the velocity vector for a cell in the flow small enough so that the velocities of the particles in the cell are approximately equal.}
\end{itemize}

Note that the assumption that the flow consists of these cells amounts to the fluid approximation.  In the particle view the existence of these cells is not assumed and the macroscopic fluid velocity is superimposed on thermal components of the particles` velocities.  When the flow is incompressible and steady, the {\it energy density} is conserved in the flow.
 
Bernoulli\textquotesingle s equation is an expression of the conservation of energy, a checksum that is very useful in the calculation of the properties of a steady flow.  It does not speak to the question of cause and effect however.  Fluid flow is caused by a pressure differential, inertia of the massive particles already flowing or direct collisions of the fluid particles with surfaces, e.g., a propellor.  In some circumstances a flow can also give rise to a pressure differential, the cause of the Coand{\v a} effect.  Just because a fluid is flowing does not mean that the pressure within the fluid has decreased, except in cases where Bernoulli\textquotesingle s equation holds.  Velocity is relative to the inertial frame where it is measured but pressure is a quantity independent of the inertial frame.  The pressure in a fluid is measured as the momentum transfer of the fluid particles striking some transducer that produces a pointer reading.  The force that moves the pointer is the integral over the (oriented) surface area of the transducer of the rate the fluid particles transfer momentum to it.  

\begin{equation}\label{eq:pr}
Pointer~Reading \propto {\mathbf Force} = m \times{\sum_{
\begin{array} {c} 
transducer \\
surface
\end{array}}}
{\frac {d{\mathbf v}} {dt}} \otimes d{\mathbf A},\end{equation}

where $m$ and ${\mathbf v}$ are the particle\textquotesingle s mass and velocity and $d{\mathbf A}$ is an area element.   We assume that the particle collisions with the surface are perfectly elastic, so the tensor product, $\otimes,$ gives a result normal to the surface element, $d{\mathbf A}$.

\begin{figure}[!htb] 
  \centering
     \includegraphics[width=80mm]{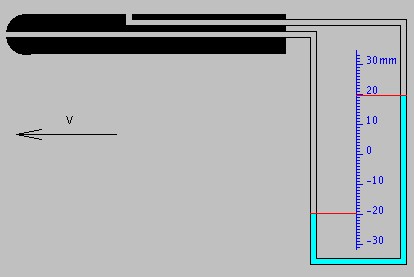}
    \caption{Pitot tube}
     \label{fig:PitotTube}
\end{figure}

The orientation of the transducer surface in the flow affects the pressure reading.\footnote{A Michaelson interferometer with a vacuum chamber in one leg can be used to measure air density from which the pressure can be calculated from thermodynamic principles. It does not measure pressure directly however.} The tensor product between the area tensor, ${\mathbf A}$ and the particle velocity ${\mathbf v}$ in Equation \ref{eq:pr} is a force which the transducer converts to a pointer reading.   A careful examination of a common Pitot tube used to measure the speed of an airplane (Figure \ref{fig:PitotTube}) will show that the speed is measured as the difference in pressure between pressure sensor areas that are normal to one another in the same flow.  (In the figure, $V$ is the velocity of the aircraft.)

A Pitot tube is a device to measure air speed, i.e., the velocity of the tube with respect to the local ambient air.  If the tube is correctly mounted on an aircraft flying in air that is not moving with respect to the earth, it measures, after altitude correction, the ground speed of the aircraft.  Its design makes use of Bernoulli\textquotesingle s relation.  It actually consists of two concentric tubes.  The outer tube is welded to the rim of the inner tube at one end and both tubes are sealed at the other end except for a manometer or other relative pressure gauge which is connected between the two tubes.  A close examination of one design of a Pitot tube will reveal small holes in the side of the exterior tube.  These holes are exposed to the air flow.  In order for the device to work correctly, it is very important that the surfaces of these holes be parallel to the flow so there is no ram pressure or rarefaction of the air there.  It is the pressure in the outer tube that is compared to the ram pressure in the center tube.  This pressure remains at ambient no matter what the air speed, even zero.\footnote{Ambient pressure is a function of altitude and so a correction must be made to the Pitot tube reading.}  It is the ram pressure in the inner tube that changes as the airspeed changes.    

Although Bernoulli\textquotesingle s equation employs densities as factors in the potential and kinetic energy terms, the equation in this form is only valid when the fluid can be assumed incompressible and non-viscous because compression heating and viscous interactions create heat energy.  To account for this energy, thermodynamics would have to enter the equation and a thermodynamic process be identified.  This process could vary in many different ways, depending in detail on the specific case.  It is for this reason that there is no heating term in Bernoulli\textquotesingle s equation.  If compression is significant, Bernoulli\textquotesingle s equation in this form cannot be expected to hold.\footnote{See http://www.efunda.com/formulae/fluids/bernoulli.cfm\ }

\subsection{Bernoulli at the particle level}
Strictly speaking, Bernoulli\textquotesingle s equation does not apply over a real free surface because particles will move lateral to the flow after striking protuberances on the surface, violating a Bernoulli assumption.\footnote{If the diameter of a real tube is much greater than the size of the wall\textquotesingle s microscopic protuberances, the tube is a Bernoulli tube to a good approximation, however.}  

Think of a pressure vessel of a non-viscous gas feeding a Bernoulli tube (a real one, glass).  Before flow starts, the energy in the vessel is equally distributed between the 3 degrees of freedom.  When the fluid is vented into the tube, the pressure in the tube is less than that in the vessel.  If the venting is sudden, a pressure wave will travel in the tube at the speed of sound and thus will precede the air flow.

The reason that the pressure in the exit tube is less than in the vessel is that the only particles that exit into the tube are those with velocity components in the exit direction.  Of course these particles exert a transverse pressure lower than that of the vessel since they are selected for their momentum components being outward into the tube.  Because energy is conserved, these particles` initial energy density is now apportioned between pressure on the walls of the tube (the pressure read by manometer) and the kinetic energy density of their velocity in the tube, ${\frac 1  2} {\rho v^2}$.   This means that there will be a lower manometer reading in the exit pipe than in the vessel.  The pressure difference between the vessel and the end of the exit pipe allows the flow of the exiting particles.  At the particle level, Bernoulli\textquotesingle s equation, where the exit tube is in the x-direction, is:

\begin{equation}
\label{eq:pvessel}
p_{\it vessel} = {\frac 1  2} \rho\sum(v_{\it x}^2 + v_{\it y}^2 + v_{\it z}^2) = p_{\it tube} + {\frac1  2} \rho \sum {v_{\it x}^2} ,
\end{equation}

where the sums are over the velocities of the all particles in the flow and

\begin{equation} 
\label{eq:pwall}
p_{\it tube} = {\frac 1  2} \rho\sum(v_{\it y}^2 + v_{\it  z}^2)
\end{equation}

 is the pressure at the tube wall.
 
At the exit orifice, it is just those particles that are moving toward the hole that actually exit.  The hole is a sorting mechanism hence the entropy decreases in the exit flow.  This sorting process at the exit selects particles that will give a lower pressure when that pressure is measured at an orifice whose plane is parallel to the flow, such as a manometer connection.

\subsection{Venturi\textquotesingle s tube}

\begin{figure}[!htb] 
  \centering
     \includegraphics[width=80mm]{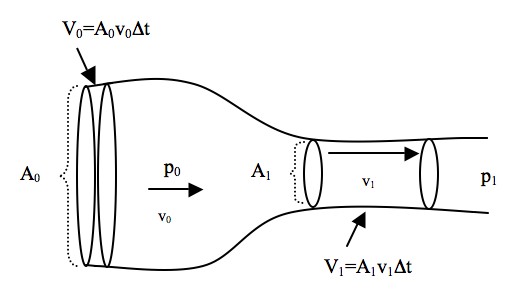}
    \caption{Venturi tube}
     \label{fig:VenturiTube}
\end{figure}

Consider a level $(\Delta h = 0)$ Venturi tube (Figure \ref{fig:VenturiTube}) connected between two large pressure chambers, one at pressure $p_0$ and the other at pressure $p_1$ where $p_1 < p_0$.  The cross-sectional area of the tube varies from some $A_0$ to a smaller area, $A_1$ in the throat.  The flow velocities are $v_0$ and $v_1$.   Assume that both diameters are much larger than the microscopic roughness of the tube wall.  Say further that the fluid flow is isothermal and inviscid, i.e. steady, and that all collisions, particle-particle and particle-wall are perfectly elastic.  This means that Bernoulli\textquotesingle s equation holds approximately, i.e., energy density is conserved in the flow and the volumes $V_1$  and  $V_0$ are equal since the mass flow rate is conserved.

What does this mean at the particle level?  A manometer reading is caused by the transfer of momentum of particles impinging on its transducer, i.e. a column of liquid, a diaphragm or some other object whose reaction is converted to a pointer reading.  These recording devices convert the transfer of the particles`  transverse momenta to a force normal to the transducing surface.  

When an orifice is opened in a pressure vessel, it sorts out and allows to exit those particles which are at the orifice and which have velocity components in the direction of the plane of the orifice.  The components of the exiting particles` velocities normal to this direction will necessarily be smaller than those of particles which do not exit.  (See Equation \ref{eq:pvessel})  If a manometer is fitted to the wall of the tube, the transverse pressure can be measured at the entrance.  As the tube\textquotesingle s diameter decreases, there is a further sorting process so that the pressure in that section is lower still.  Particles in the tube that are outside the imaginary projection of the narrow tube back into the larger section, will strike the curving wall of the neck and interact with other particles.  They bounce off elastically with undiminished energy and with a change of momentum. They will then energize the particles near the small-diameter exit tube.   The result of these interactions is the conservation of energy and the transfer of the energy in the annulus to the particles in the smaller tube.  

Rewriting Equation \ref{eq:enden} with $h=0$ and adding some more detail, we get

\begin{equation}
\label{eq:VentPart}
{\frac 1  2} {\frac m  {V_0}} \sum(v_x^2+v_y^2+v_z^2)_0 = {\frac 1  2} {\frac m  {V_1}} \sum(v_x^2+v_y^2+v_z^2)_1,
\end{equation},

where the sums are over the particles in $V_0$ and $V_1$ respectively.  With the main fluid velocity in the $x-$direction, the conservation of mass yields

\begin{equation}
\label{eq:ConsMass}
\left(\bar{v}_x\right)_0 = {\frac {A_1}  {A_0}}\left( \bar{v}_x\right)_1.
\end{equation}

where $\bar{v}$ is the average velocity component.

Further, since $V_0 = V_1 = V$ we replace $\sum v^2$ by $N\bar{v}^2$ where $N$ is the number of particles in the volumes $V_0$ and $V_1$, and put $\rho =  {\frac {mN}  V}$.   The pressures measured by manometers in $V_0$ and $V_1$ are, respectively, $p _0 =  {\frac 1  2}\rho(\bar{v}_y^2 + \bar{v}_z^2)_0$ and $p _1 =  {\frac 1  2}\rho(\bar{v}_y^2 + \bar{v}_z^2)_1$ so, with some algebra, we have the Venturi relation,

\begin{equation}
\label{eq:Venturi}
\left(\bar{v}_x\right)_1 = \sqrt{ {\frac {2\left(p _0 - p _1\right)}  {\rho\left[1 - \left({\frac {A_1}  {A_0}}\right)^2\right]}}}.
\end{equation}

It is clear from this development, then, that the higher velocity in the Venturi throat is not the cause of the lower pressure there.  The lower pressure and the higher velocity are both due to the sorting effect of the narrowing tube and the complex interactions of the particles as they enter the throat.

There is an interesting result of Bernoulli\textquotesingle s equation in the form of Equation \ref{eq:VentPart} and a result of Statistical Mechanics.  According to Statistical Mechanics, the root mean square of the molecular speed, $v_{rms}$ is related only to the temperature, $T$ and the molecular mass, not to the pressure, $p$.

\begin{equation}
\label{eq:Tempeq}
{\frac 1  2} mv^2_{rms} = {\frac 3  2} k_B T,
\end{equation}

where $k_B$ is Boltzmann\textquotesingle s constant, and thus no matter what $p_0$ is, $\left(\bar{v}_x\right)_0$ {\em cannot exceed the molecular speed, $v_{rms},$ corresponding to the temperature $T$!}  This is why it is important and fortunate that the high pressure in a rocket motor is created as an effect of high temperature.  The beneficial effect of high temperature in rocket propulsion will be seen below in Section \ref{REdiffuser}.

By the same token the temperature of the gas exiting the rocket\textquotesingle s plenum is cooler than that in the plenum because some of its energy is converted from heat to the energy of translation. In a reference frame moving with the exiting gas $v_{rms}^2$ will have dropped at the expense of the velocity of the moving frame and so, according to Equation \ref{eq:Tempeq}, the temperature of the exiting gas will be lower than it was in the plenum.  In the plenum, the effect of this more or less adiabatic process is to rapidly reduce the pressure there and thus, since $p = k_B \rho T,$ and assuming that the plenum is large enough so that the percent plenum depletion rate is much less than the exit rate of the gas, i.e. $\rho$ is approximately constant in the plenum, the temperature in the plenum decreases too.  

\begin{center} ****************** \end{center}
There is a rather new device that takes advantage of this sorting effect. It is The Dyson Air Multiplier\textsuperscript{\texttrademark} (Section \ref{DysonCool}). While a conventional cooling fan blows the warm room air and actually adds energy, the cooling it provides is mainly only due at to evaporation of perspiration, the Dyson device uses an internal turbine to pressurize a plenum.  Of course the action of the internal compression heats the air in the plenum, but the air exiting the plenum is forced out a narrow annulus and the sorting described above occurs.  The result is a breeze whose coolness can be easily felt before that air comes into equilibrium with the ambient air in the room.  The non-equilibrium maintained by the device results in a pleasant cooling effect without a refrigeration unit.

\subsubsection{The two-fluid atomizer}

\begin{figure}[!htb] 
  \centering
     \includegraphics[width=80mm]{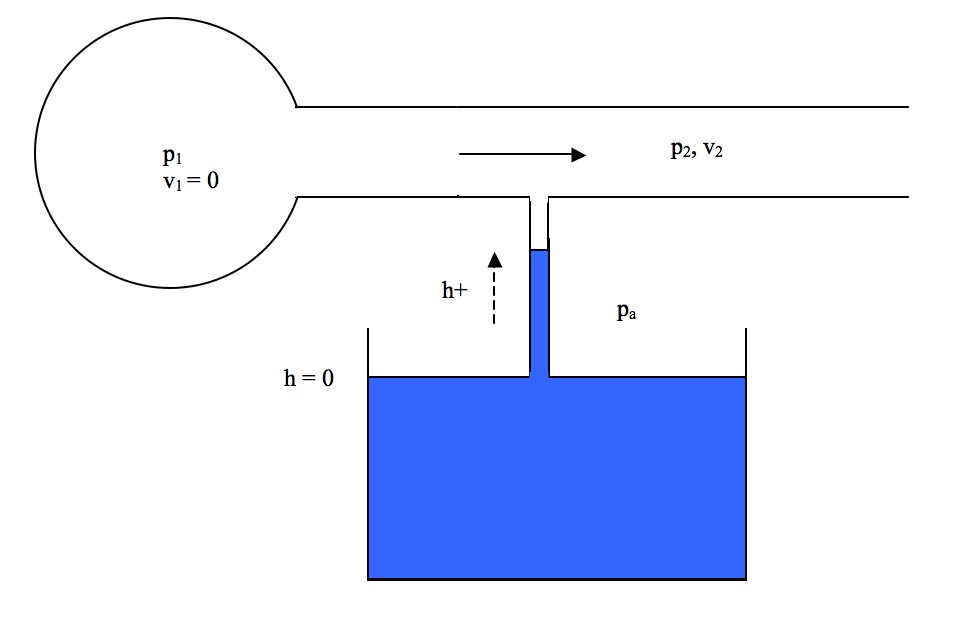}
    \caption{Theoretical atomizer}
     \label{fig:Theoretical_atomizer}
\end{figure}

Atomizers are often cited as examples of devices that make use of the Bernoulli principle.  Figure \ref{fig:Theoretical_atomizer} is an illustration of what is commonly understood to be a two-fluid atomizer.

The vessel on the left represents a flexible bulb filled with air.  The pressure $p_1$ is created when someone squeezes the bulb.   Although the molecules of air are moving in the bulb, their motion is random in every direction so, at the macroscopic level and under the fluid approximation, the velocity $v_1 = 0$. When the air is flowing steadily in the horizontal tube, however, pressure $p_2$ and velocity $v_2$ develop.
  
The vertical tube is connected to the atomizer and its lower part is submerged in the liquid meant to be atomized and mixed with the air.  The air velocity in the vertical tube is zero.  Finally the mixture is expelled into the atmosphere at pressure $p_a$.  Below are the parameters in the regions of the apparatus:

\begin{itemize}
\renewcommand{\labelitemi}{$\ $}
\item{$p_1$ is the pressure in the bulb,}
\item{$p_2$ and $v_2$ obtain in the horizontal tube,}
\item{$p_a$ is the atmospheric pressure and}
\item{$h$ is the liquid height in the vertical tube.}
\end{itemize} 

If we assume that the air density, $\rho$, is constant then Bernoulli\textquotesingle s equations become

$$p_1 = p_2 + {\frac 1 2} \rho v_2^2 $$ and $$p_2 + \rho_w g h = p_a,$$

where $\rho_w$ is the density of the liquid in the reservoir.

These relations result in

\begin{equation}
\label{eq:BAtomizer}
p_{gauge} = {\frac 1 {\rho v_2^2 - \rho_w g h}}
\end{equation}

where $p_{gauge} = p_1 - p_a$ is the gauge pressure in the bulb and $g$ is the acceleration due to gravity.
Equation \ref{eq:BAtomizer} shows that the height of the liquid is a function of both the air velocity, $v_2$, and the gauge pressure, $p_{gauge}$, in the bulb. Of course,  $v_2$ is a function of $p_{gauge}$. This relation also predicts that if the bulb is squeezed suddenly, the height, $h$, will be negative.  The pressure pulse travels a the speed of sound but the atomizer can only function correctly when $v_2$ has had time to build up.  \subsubsection{Conventional atomizer}
\begin{figure}[!htb] 
  \centering
     \includegraphics[width=70mm]{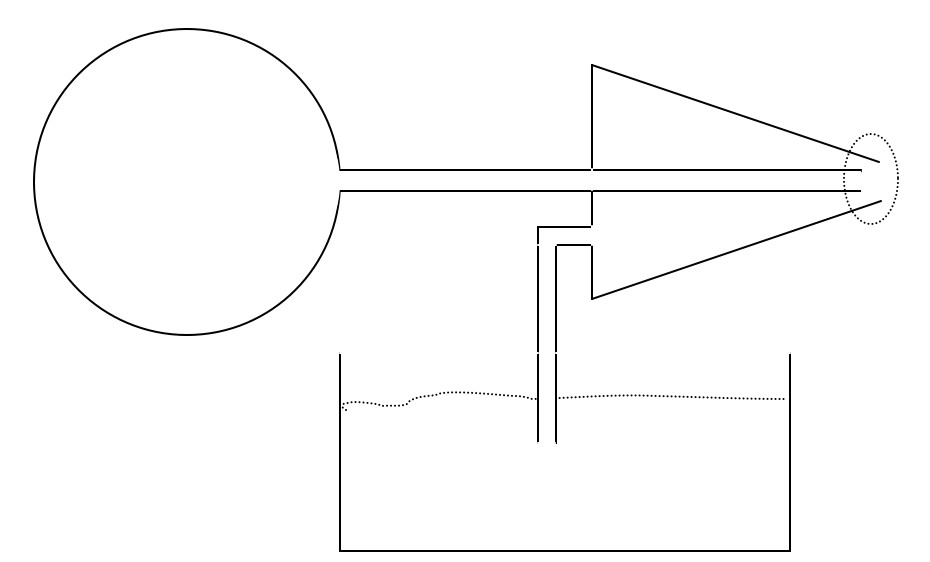}
    \caption{Real atomizer}
     \label{fig:Real_atomizer}
\end{figure}

In many physics books Figure \ref{fig:Theoretical_atomizer}, or its equivalent, is displayed as a schematic of an actual two-fluid atomizer.  A search of patent disclosures for two-fluid atomizers reveals, however, that none of them is designed like this.  Figure \ref{fig:Real_atomizer} is a diagram of a conventional two-fluid atomizer.

\begin{figure}[!htb] 
  \centering
     \includegraphics[width=70mm]{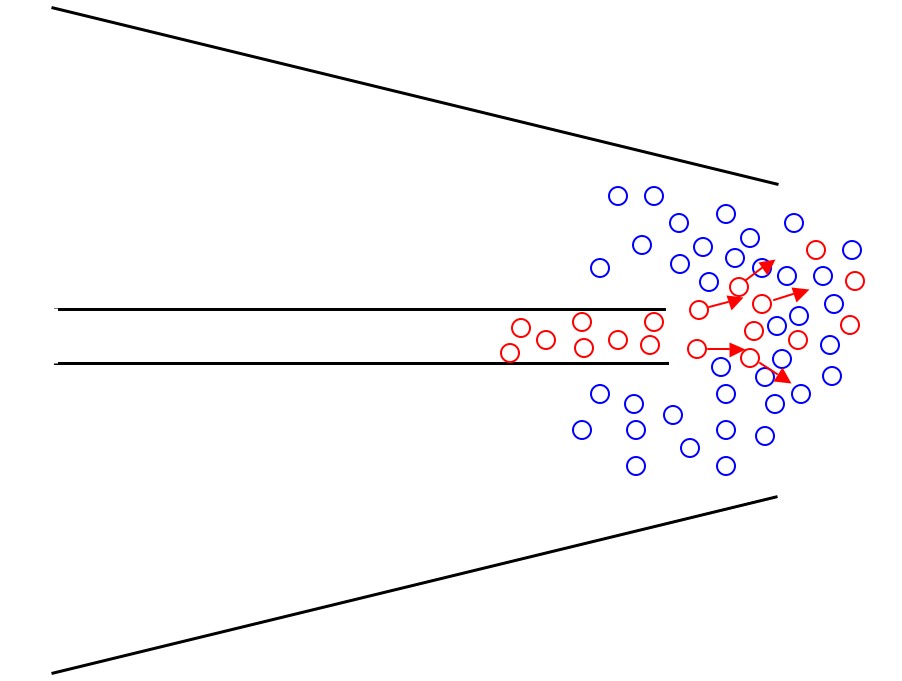}
    \caption{Nozzle Detail}
     \label{fig:Nozzle_Detail}
\end{figure}

The pressure is applied from the ball on the left.  High-pressure air exits the inner tube on the right and sweeps first the air from the conical volume and then the fluid from the chamber connected to the reservoir as it is ``drawn up`` by the low pressure created in the dotted area.  This process is detailed in Figure \ref{fig:Nozzle_Detail}.

The particles with arrows exit the high pressure tube and collide directly with the molecules of the other fluid, mixing with them and sweeping them away and out of the nozzle.  In order to achieve the best mixing and delivery, the detailed design of this region is very important.

It is also important to note that, in this mixing and ejection region, Bernoulli\textquotesingle s assumptions do not hold so Bernoulli\textquotesingle s equation cannot be used.  Recall that Bernoulli\textquotesingle s equation assumes that there is {\it no interaction} between the fluids nor with the surfaces in the flow.  In the case of a two-fluid atomizer, however, mixing and delivery are the two important goals. 

\subsubsection{Flit gun}
 
\begin{figure}[!htb] 
  \centering
     \includegraphics[width=80mm]{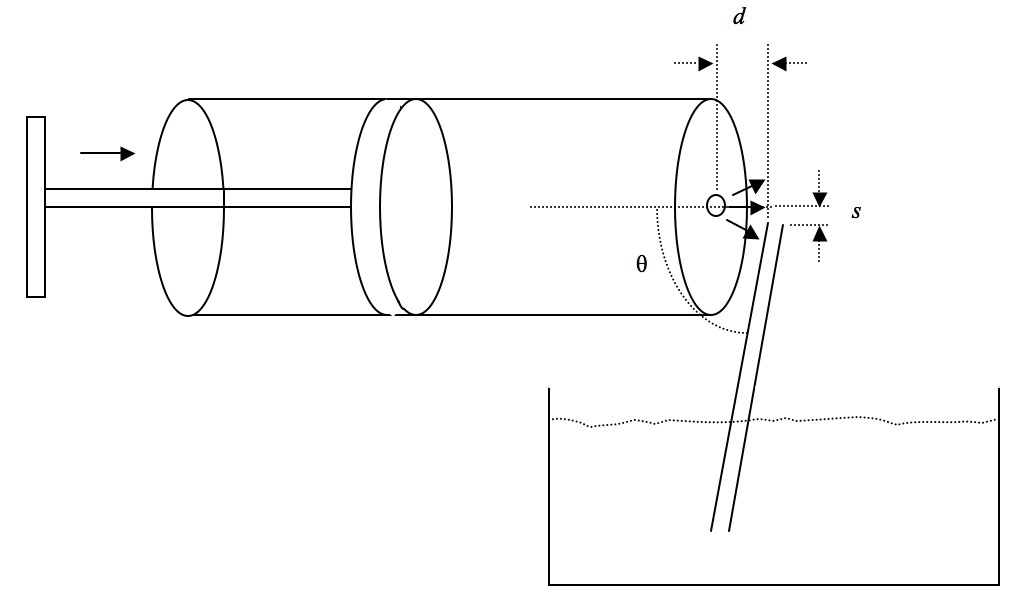}
    \caption{Flit Gun}
     \label{fig:Flit_Gun}
\end{figure}

Figure \ref{fig:Flit_Gun} is a diagram of the famous ``Flit Gun`` bug sprayer.
 
The air stream exiting the gun passes over the tube and interacts with the fluid in it.  First a partial vacuum is created in the tube then the liquid rises in the tube and mixes with the air exiting the pump.  The angle, $\theta< 90^\circ$,  is crucial as are the distances, $s$ and $d$.  If, for example, $\theta \ge 90^\circ$, the liquid will be forced back into the reservoir.  Bernoulli\textquotesingle s equation does not hold for this extremely unsteady flow.

\subsubsection{Flow into an expansion chamber}

\begin{figure}[!htb] 
    \centering
     \includegraphics[width=80mm]{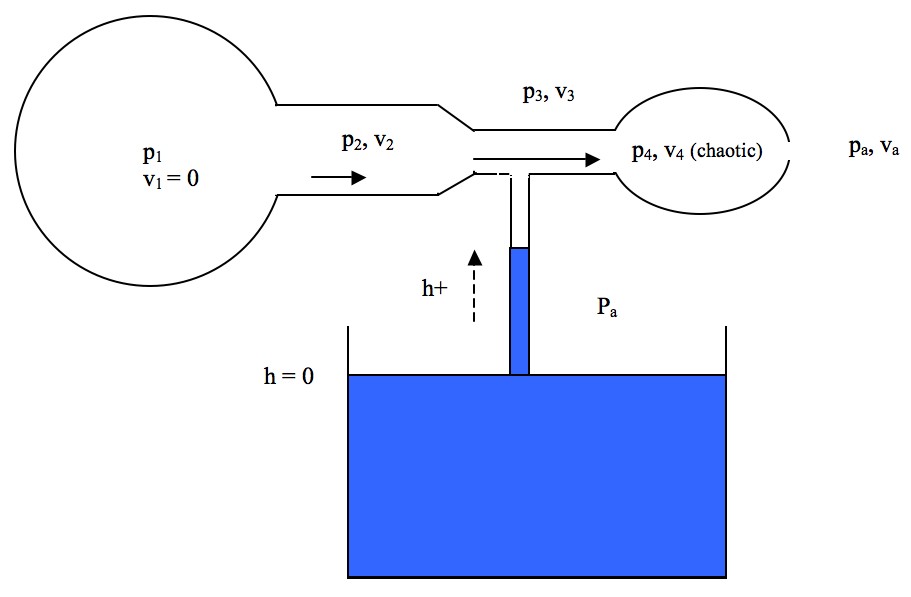}
    \caption{Restricted exit orifice}
     \label{fig:Restricted_exit_orifice}
\end{figure}

We refer to the sections in Figure \ref{fig:Restricted_exit_orifice} as Sections $1, 2, 3, 4$ and $a,$ the region under atmospheric pressure. 
The figure shows an apparatus in which high-velocity air from a low pressure region flows into a chamber at higher pressure. If region $p_4$ were to be sealed off completely, the flow into it from region $3$ would be reminiscent of Maxwell\textquotesingle s displacement current, the current flowing into a discharged capacitor. As the capacitor builds up charge, the displacement current decreases.
 
It is important to remember Equation \ref{eq:pvessel} when thinking about the pressures in the different sections.  Bernoulli\textquotesingle s equation, Equation \ref{eq:enden}, does not rule out $p_3  <  p_4$ in Figure \ref{fig:Restricted_exit_orifice}.  In fact that happens when the pressure, $p_1$, is high enough.  It may seem counterintuitive that the fluid can flow from a lower pressure into a higher one but the forward momentum of the high velocity particles incoming from Section $3$ is greater than that of those already in Section $4,$ i.e.,
$$(m\overline{v_x})_3 > (m\overline{v_x})_4,$$
and thus the particles entering win in the contest of collisions with the particles in Section $4$ and actually enter that section.  In contradistinction to the situation upstream where the pressures are decreasing and the interparticle interaction is negligible, there are now many collisions between particles.  In the process, the energy of forward motion of the entering gas is in part converted into internal energy of the gas, i.e. heat, pressure and molecular energy of vibration and rotation, depending on the structure of the gas molecules.  We  can call this {\em partial thermalization} because some of the kinetic energy of the molecules in translation into Section $4$ is converted to heat and internal molecular excitation, while some appears as macroscopic eddies and turbulence.

Later, in Section \ref{REdiffuser} of this paper, we will see how the temperature of the fluid flowing into a region like Section $4$ of Figure \ref{fig:Restricted_exit_orifice} but with an open end to the atmosphere can cause a thrust enhancement in a rocket engine.

\subsection{The Joule-Thomson effect}

The above situation lies somewhere between the case of Bernoulli flow, i.e. no particle interaction in a smoothly flowing fluid, and the behavior of a real, self-interacting gas which produces the Joule-Thomson\cite{JT} effect.  In this latter regime the effects of particle collisions are of paramount importance.

 Instead of Equation \ref{eq:enden}, we will use the form below for the specific energy density as the conservation law.
 
\begin{equation}
\label{eq:ConsEnthalpy}
{\frac 1 2}v^2 + w = constant
\end{equation}

where $w$ is the {\em enthalpy,} 

\begin{equation}
\label{eq:Enthalpy}
w = \epsilon + p/\rho
\end{equation}

and $\epsilon$ is the {\em thermodynamic energy} per unit mass of the gas.  This is the energy of interatomic oscillation (in case of polyatomic gases) and rotation of the molecules as well as their potential energy due to the van der Waals forces between them.

\begin{figure}[!htb] 
    \centering
     \includegraphics[width=70mm]{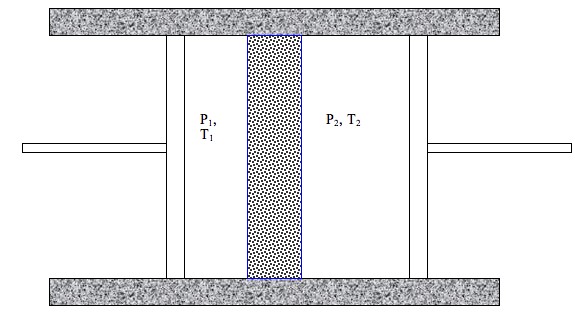}
    \caption{Joule-Thomson apparatus}
     \label{fig:JTeffect}
\end{figure}

In the case that the flow velocity vanishes, $v=0,$ equation \ref{eq:ConsEnthalpy} expresses the conservation of Enthalpy.  Such a case is obtained in the apparatus shown in Figure \ref{fig:JTeffect}.

In order that the thermodynamic process be adiabatic, i.e. no heat flowing in or out, the apparatus is insulated.  The piston on the left provides pressure, $P_1$ at temperature $T_1,$ and the gas moves to the chamber on the right, at pressure $P_2$ and temperature $T_2$ by passing through a so-called ``porous plug`` made of packed granules of a chemically non-interacting substance, for example frit made of silica.  As the gas molecules bounce in chaotic fashion against each other and the frit particles, the flow velocity is transformed into random motion and thermodynamic energy.  The following quote is taken from the Wikipedia article on the Joule-Thomson effect:

\begin{quotation}
\small{\noindent ``As a gas expands, the average distance between molecules grows. Because of intermolecular attractive forces (Van der Waals forces), expansion causes an increase in the potential energy of the gas. If no external work is extracted in the process and no heat is transferred, the total energy of the gas remains the same because of the conservation of energy. The increase in potential energy thus implies a decrease in kinetic energy and therefore in temperature.

A second mechanism has the opposite effect. During gas molecule collisions, kinetic energy is temporarily converted into potential energy. As the average intermolecular distance increases, there is a drop in the number of collisions per time unit, which causes a decrease in average potential energy. Again, total energy is conserved, so this leads to an increase in kinetic energy (temperature). Below the Joule-Thomson inversion temperature, the former effect (work done internally against intermolecular attractive forces) dominates, and free expansion causes a decrease in temperature. Above the inversion temperature, gas molecules move faster and so collide more often, and the latter effect (reduced collisions causing a decrease in the average potential energy) dominates: Joule-Thomson expansion causes a temperature increase.``\cite{JT}}
\end{quotation}

A device called a ``heat pump,`` using an appropriate gas, can be a heater or cooler depending on how the pressures $P_1$ and $P_2$ are adjusted. 

\section{The Coand\v{a} effect} \label{sec:coanda_effect}
This effect, first investigated and employed by the Romanian aerodynamics engineer Henri Marie Coand\v{a} (1886 -- 1972), usually refers to the phenomenon in which an air flow attaches to an adjacent wall which curves away from this flow. (see \cite{schlichting} pp. 42, 664).  Another aspect of this phenomenon is the entrainment of molecules far from the jet (see the three panels in \ref{fig:WingFlow}).  The attachment effect is taken for granted and it is the {\it separation} of the flow from an aerodynamic body that is discussed as a precursor to the stalling of the surface (\cite{schlichting} p. 40).  The effect can be seen in some automobile advertisements.  Streamers of smoke are seen to hug the profile of a car in a wind tunnel even as the surface of the car curves away from the flow.  This behavior indicates a lower pressure that aerodynamicists call {\em suction} at that part of the surface.  This is puzzling since there are no long-distance attractive forces acting in a gas or between the gas molecules and a surface under normal\footnote{A fluid in liquid form behaves differently.} conditions.  Figure \ref{fig:CurvedSurface} in section \ref{Flfl} discusses the mechanism for suction.

\subsection{Organ pipe beard}
The Coand\v{a} effect is exploited in the design of large flue pipes in some pipe organs.\footnote{Organ builder Bill Visscher, private communication.}  These pipes are like huge whistles and can, if they are overblown, sound the octave rather than the fundamental tone.  Anyone who has played an Irish tinwhistle knows this effect.  Much of the awesome power of the grand organ, however, comes from the volume of the bass notes.  The pipe sounds when a sheet of air is blown over the mouth.  Some of this air enters the pipe and of course it must also exit.  The only exit from these closed pipes is the mouth itself.  The exit path, then, starts at the top of the mouth of the pipe.  The unwanted octave sounds when air exiting from the mouth interferes with the wind sheet entering the pipe.  How, then, to avoid this interference?

\begin{figure}[!htb] 
  \centering
     \includegraphics[width=70mm]{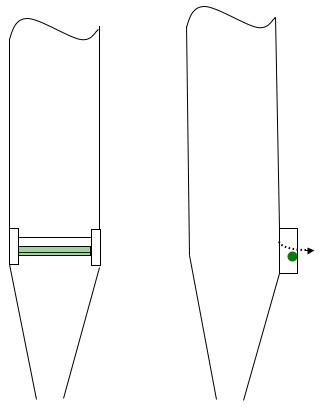}
    \caption{Beard on an organ flue pipe}
     \label{fig:beard}
\end{figure}

Some organs utilize what are called {\em beards} to direct the air flowing out of the pipe away from the air entering from the air chest.  A beard is a circular dowel mounted between the ears on each side of the mouth.  As the air exits, it tends to flow in the general direction of the beard but the beard is located so that the main flow passes {\em over} it.  As the surface of the beard curves away from the flow, a low pressure is created on the top of the beard.  This low pressure area attracts the flow and keeps it from interfering with the flow entering the pipe.  Figure \ref{fig:beard} illustrates this.  The precise location and size of the beard also affect the timbre of the pipe\textquotesingle s sound.

\newpage
\subsection{The Bunsen burner}
A common application of the effect illustrated in Figure \ref{fig:WingFlow} is the Bunsen burner.  An exploded view is shown in Figure \ref{fig:BunsenBurner} below.

\begin{figure}[!htb] 
  \centering
     \includegraphics[width=70mm]{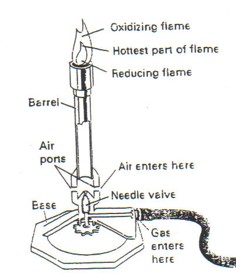}
    \caption{Bunsen Burner}
     \label{fig:BunsenBurner}
\end{figure}

The effect shown in the first panel of Figure \ref{fig:WingFlow} is utilized in the Bunsen burner and any other burner where the fuel and air is mixed upstream of the flame.  The needle valve regulates the rate of fuel flow and the Coand\v{a} effect accounts for the mixing as air is drawn in from outside the burner.

\subsection{The Coand\v{a} propelling device}
Henri Coand\v{a} held many patents but perhaps the most interesting for aerodynamic design is his design of a propelling device \cite{coandapatent}.  The device develops lift as an enhanced buoyant force produced by decreased pressure on the top.  This decreased pressure on the curved circular surface is caused by a flow of air at high-pressure exhausting tangentially to an annular airfoil from an annular slit.  In addition to enhancing the buoyant force, the device removes the bow wave that would otherwise hinder the motion of the device.  

A bow wave is normally formed when an object moves through a fluid.  It is easy to see the bow wave of a ship or barge.  As it is propelled in the water, a ship must push water out of the way.  Because the water has mass, force is required to move it.  By Newton\textquotesingle s third law, there is an equal and opposite force exerted on the ship.  This effect causes drag in addition to the viscous drag along the hull of the ship as it moves through the water.  

A toy helium balloon rises much more slowly than if it weren`t hindered by a bow wave in the air.  It is primarily the force of the bow wave that is responsible for the phenomenon of {\em terminal velocity}.  By extending his arms and legs, a skydiver can control his terminal velocity, increasing or decreasing it, by manipulating the area of the surface presented to the air. It is the bow wave of the total surface that causes the third law force impeding the fall.  

Figures \ref{fig:Coandaproptotal} and \ref{fig:Coandaprop} below illustrate the function and also the dissipation of the bow wave by the Coand\v{a} propelling device.

\begin{figure}[!htb] 
  \centering
     \includegraphics[width=70mm]{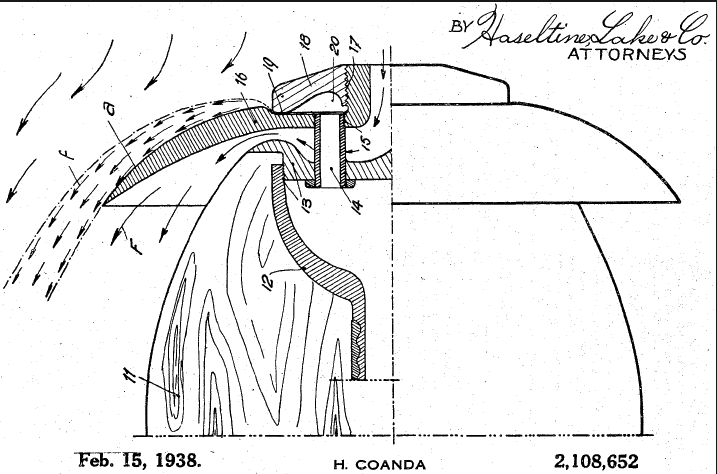}
    \caption{The Coand\v{a} Propelling Device}
     \label{fig:Coandaproptotal}
\end{figure}

The detail in Figure \ref{fig:Coandaprop} below shows its function.

\begin{figure}[!htb] 
  \centering
     \includegraphics[width=70mm]{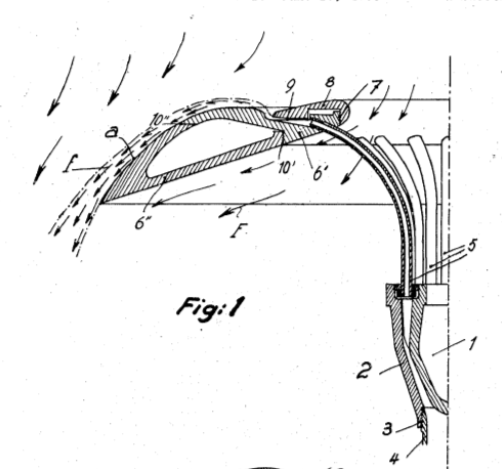}
    \caption{Detail of the Coand\v{a} Propelling Device}
     \label{fig:Coandaprop}
\end{figure}

High velocity air is pumped through pipes (5) and exits over the top of the device through the orifices (9).  This air exhaust (f) flows over and adheres to the surface (a).  The difference between ambient pressure and the lower pressure on (a) causes the ambient air to be drawn to it.  This is indicated by the arrows over the surface (a) and the arrows near (7), the opening at the top of the device. At (a) it can be seen that the bow wave is swept away.

\chapter{Calculation of lift}\label{lift_calc}
Lift is caused by the collisions of fluid particles with the surface of the airfoil.  By Newton\textquotesingle s third law, this interaction of the particles with the surface results in an equal and opposite reaction on the airflow itself; the particles bounce back.  Say, for example, that the lift force is in the ``up" direction, then the third law force on the air is ``down."\footnote{Here, we are only concerned with the lift force on heavier than air craft.  The situation is different for aerostats.}    The lift can be represented in two ways: 1) as the summation of all the forces on the surface or, according to Newton\textquotesingle s third law, 2) by the negative of the force the surface exerts on the air.  The latter is the approach that led to the Kutta-Joukowski theorem.\footnote{See Reference \cite{anderson} pages 236 and 237.}

\begin{figure}[!htb] 
  \centering
      \includegraphics[width=70mm]{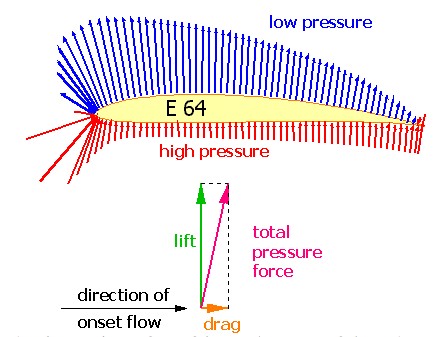}
      \caption{Typical force configuration on an airfoil in an air flow}
      \label{fig:forces_hepperle}
\end{figure}

Figure \ref{fig:forces_hepperle} shows a typical force configuration on the surface of an airfoil.\cite{hepperle}  The air flow is from the left.  Note that the primary contribution to the lift is from the curved surface of the {\em top} of the wing.  This lowered pressure, the so-called {\em suction,} created there also causes the Coand\v{a} effect.  

The flow particles far from the airfoil\textquotesingle s surface ``feel``  this suction as a sort of reverse bow wave and, as a result, flow toward the surface.  The low pressure, maintained by the flow past the curved surface, results in a pressure gradient, $\partial p /  \partial \xi,$ that decreases to zero as $\xi$, the normal distance from the airfoil, increases. The pressure approaches the limit $p_\infty,$ the ambient pressure. 

Another interesting aspect of this figure is the indication of a (conventional) bow wave of positive pressure just below the leading edge.  This bow wave results in an upwash that moves against the main flow to join the flow above the leading edge stagnation point.  At high angles of attack this flow causes a vortex on the top of the wing which becomes larger as the angle of attack and/or the flow velocity increases.  (See the third panel in Figure \ref{fig:WingFlow} above.) This vortex interferes with the suction on top of the wing and if too large will eventually cause stall.  Vortex generators, sometimes mounted on wings and control surfaces,\cite{barrett} in spite of their name, inhibit the formation of this span-wise vortex.  They do this by generating small vortices emanating from their tips. These small vortices, for angles of attack not too large, break up the larger span-wise vortex before it forms.  The axes of these vortices are in the direction of the flow.

A common stall warning device is a switch activated by a simple flap mounted on the leading edge and protruding forward, which, when it gets blown upward, causes a horn in the cockpit to sound.  All stall warning devices are activated, directly or indirectly by the speed of the upwash.\cite{denker}  Upwash is created by the viscous interaction of the air with the lower surface of the wing.  It can be seen as a stream of water from a faucet strikes a plate held at an angle to the stream.  Some of the water flows upward before it finally turns and flows down the plate.  If the plate is held so the stream is near the top, the upwash will actually run up and over the top of the plate.  As we will see in Sections \ref{slots} and \ref{birds}, this upwash can be turned to advantage to increase lift.

The total force on the wing is the sum of lift and drag (the red arrow in Figure~\ref{fig:forces_hepperle}). This force is the vector sum:

\begin{equation*}
\tag{\ref{eq:Ftotal}}
{\mathbf F}_{total} =  -\oint_{airfoil} \left({d {\mathbf F}_n +  d {\mathbf F}_s}\right).
\end{equation*}

where

 \begin{itemize}
\renewcommand{\labelitemi}{$\ $}
\item{${\mathbf  F}_n$ is the force normal to the surface and}
\item{${\mathbf  F}_s$ is the force tangent to the surface.}
\end{itemize} 

The minus sign is necessary because we are calculating force on the air and use Newton\textquotesingle s third law to relate that to the total force, ${\mathbf  F}_{total},$ on the surface.

Newton\textquotesingle s second law is:

\begin{equation}
\label{eq:NII}
{\mathbf  F} = m{\mathbf  a},
\end{equation}
 
 or, componentwise,
 
 \begin{equation}
 \label{eq:NIIcomp}
F_i = ma_i, \ i=n,s.
\end{equation}

$i = s$ denotes the component of the force and resulting acceleration along the surface and $i = n$ denotes the component normal to the surface.

\subsection{Using Newton\textquotesingle s Third Law: Effects on the air caused by the presence of the airfoil}

 At the surface of the airfoil, the pressure exerts a force equal in magnitude and opposite in direction on the air.  This pressure affects the air out to a distance of $\Delta y,$ often many airfoil chord lengths from the surface.  Newton\textquotesingle s second law in differential form is
 
 \begin{equation}
\label{eq:N2press}
d{\mathbf  F}_{airfoil} =  - \rho {\frac {ds}  {dt}} \cdot {\frac {d {\mathbf  v}}  {ds}} dA \  dr
\end{equation}

 where
 
 \begin{itemize}
\renewcommand{\labelitemi}{$\ $}
\item{$\rho (s,r)$ is the air density in the volume $dV = ds \times dr \times unit\ span.$}
\item{${ {ds} / {dt}} = v(s,r) =\ \vline\, \!{\mathbf  v}(s,r)\ \! \vline$ is the air speed,}
\item{${\mathbf  v}(s,r)$ is the velocity of the air, }
\item{$dA$ is the differential surface area element,}
\item{$r$ is the distance normal to the surface at $ds.$}
\item{$s$ is the distance along the surface.}
\end{itemize}

Again, the minus sign is required by Newton\textquotesingle s third law since we are interested in the force on the airfoil.

The behavior of the air near the surface of the airfoil is very complex and chaotic but because at angles of attack less than the stall angle, this layer, the boundary layer, is very thin compared to $\Delta y,$ this complexity is not important.  It is similar to the behavior in the bow wave of a boat.  The water is turbulent and moving in a very complex way at the prow but some small distance away the water begins to smooth out into regular waves that fan out as the boat passes.  The information as to the detailed behavior in the boundary layer has been lost to heat due to the viscosity of the water.  The only thing that has been propagated over a long distance is the effect of the pressure in the boundary layer and even this vestige of the behavior of the air near the surface will be lost at large distances from the airfoil due to viscous heating.  It is this fact that has to be ignored in the potential flow\cite{pot_flow} approximation, e.g., the Kutta-Joukowski theorem\cite{KJtheorem}.  This theorem requires the integral called the {\em circulation}

\begin{equation}
\label{eq:circulation}
\Gamma \equiv \oint {\mathbf  F} \cdot {d {\mathbf  s}},
\end{equation}

be performed over a line encircling the airfoil but outside the boundary layer.  Far enough, indeed, so that the flow there is steady.  In a real fluid there remains no effect of the airfoil after much larger distances. The flow there is purely translational, with no rotation.
 
\begin{figure}[!htb] 
  \centering
      \includegraphics[width=70mm]{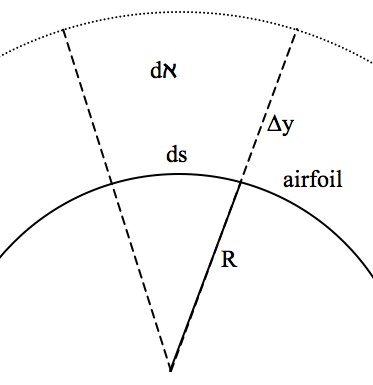}
      \caption{Geometry outside the airfoil}
      \label{fig:dvol}
\end{figure}

The presence of the surface causes shear in the boundary layer\cite{colombini} so $v \neq v_\infty,$ the flow speed far from the airfoil.  (In fact on the top of the airfoil at angle of attack, $v > v_\infty.$)  That means that the flow is not {\em steady} \cite{landau} there and Bernoulli\textquotesingle s equation does not hold.

Integrating Equation \ref{eq:N2press} there results

 \begin{equation}
\label{eq:newlift2}
{{\mathbf  F}_{per\ unit\ span} =   -\  \int_{surface}^{\Delta y} dr \oint_{C(r)} ds\ {\rho (s,r)}v(s,r){d{\mathbf  v} \over ds}.}
\end{equation}

The contour $C(r)$ follows the surface or outside the surface, the streamline contour.  

\begin{figure} 
  \centering
     \includegraphics[width=70mm]{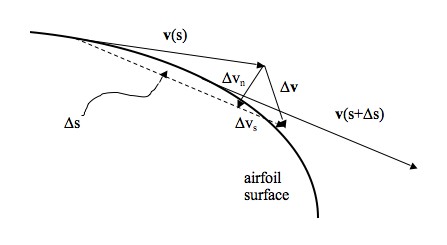}
     \caption{Illustration of the covariant derivative.}
     \label{fig:Covariant_derivative}
\end{figure}

If the flow is not separated from the airfoil, the Coand\v{a} effect, the derivative of ${\mathbf  v}$ consists of  two parts: ${\partial {\mathbf  v}} / {\partial s}$ and a geometric term that is the turning of the velocity vector due to the curvature of the airfoil.\footnote{The attached velocity field is a {\em vector bundle} over the surface of the airfoil.  This surface is assumed to be a {\em differentiable manifold.}  More information about differentiable manifolds can be found in any book on Differential Geometry.}  Figure \ref{fig:Covariant_derivative} illustrates this.

The dotted arrow in Figure \ref{fig:Covariant_derivative} is the ${\mathbf  v}(s+\Delta s, r+\Delta r)$ vector transported parallel tail-to-tail with the ${\mathbf  v}(s,r)$ vector so that $\Delta {\mathbf  v}$ can be calculated.   Taking the limit as $\Delta s \rightarrow 0,$ the {\em covariant derivative} of ${\mathbf  v}$ is obtained:

\begin{equation}
\label{eq:covder}
{\nabla_s} {\mathbf  v} = \lim _{{\Delta s} \to 0}{{\Delta {\mathbf  v}} \over {\Delta s}} = {{\partial {\mathbf  v}} \over {\partial s}} + { {\mathbf  v} \over {R(s)}},
\end{equation}

where $R(s)$ is the radius of curvature of the airfoil at $ds.$

We will now write the acceleration of the fluid at the surface as 

$${\mathbf  a} = \nabla_s {\mathbf  v} \cdot {{ds} \over {dt}}.$$

We use the covariant derivative in Equation \ref{eq:covlift} below.
 
At the surface of the airfoil and due to its presence in the flow, the pressure causes a force on the airfoil as well as on the air.   Equation \ref{eq:newlift2} then becomes
 
\begin{equation}
\label{eq:covlift}
{{\mathbf  F}_{total} =  -\  \int^{\Delta r}_{surface} dr  \oint_{C(r)} ds\ {\rho (s,r)}v(s,r){{\nabla_s{\mathbf  v}}}.}
\end{equation}

where
 
\begin{itemize}
\renewcommand{\labelitemi}{$\ $}
\item{$\rho (s,r)$ is the air density,}
\item{${\mathbf v}(s,r)$ is the velocity of the air, }
\item{$v(s,r) = {ds / dt} = \ \!\vline\ \!\!{\mathbf  v}(s,r)\ \! \vline$ is the air speed.}
\end{itemize}

The presence of the surface causes shear in the air around it\cite{colombini} so $v(s) \neq v_\infty,$  and $v_\infty$ is the flow speed far from the airfoil.  That means that the flow is not {\em steady} \cite{landau} there and Bernoulli\textquotesingle s equation does not hold.  This region is the boundary layer and its thickness is $\delta.$

Notice that the {\em circulation,}

\begin{equation*}
\tag{\ref{eq:circulation}}
\Gamma \equiv \oint {\mathbf  F} \cdot {d {\mathbf  s}},
\end{equation*}

doesn`t arise in this derivation.  In its place we have Equation \ref{eq:Ftotal}.\footnote{The circulation would be the first term in this equation, i.e. the integral of the forces normal to the airfoil.  Notice that this integral is a mathematical object and does not indicate that there is a physical vortex around the airfoil.  The resultant force on a buoyant object fixed in a quiescent fluid is written the same as the circulation integral except that the integral is over the surface of the 3-dimensional object. In this case there is no flow at all.}  We are looking at the {\em effect} on the air flow of the complex behavior of the air at the surface and in the boundary layer.  This effect exists as a reaction to the lift force.  Equation \ref{eq:covlift} replaces the Kutta-Joukowski theorem.   
 
The difficulty is in evaluating the integrals in Equation \ref{eq:covlift}.  As has been noted above, shear cannot be neglected.  The pressure, even outside the boundary layer, is not constant. The boundary layer is defined as that space, thickness $\delta,$ just outside the airfoil surface where 

$${\partial {\mathbf v}(s,r) \over {\partial r}}\ \vline_{\footnotesize 
\begin{array} {l}
 \small 
 {0 < r < \delta}
 \normalsize
\end{array} }\simeq 0$$\\

{\noindent
and ${\mathbf r}$ is in the direction normal to the airfoil.}\\

The behavior of the air in the boundary layer may be complex but for laminar flow over a non-stalling airfoil, its behavior results just in shear and a pressure gradient. The density, $\rho,$ is actually a function both of $s$ and $r.$  What value should be assigned $\rho$?  We are concerned with the {\em cause} of lift, i.e., the forces on the surface of the airfoil.  Our understanding is in terms of Newton\textquotesingle s laws.\footnote{When Isaac Newton was asked {\em why} the apple falls as it does, he is reported to have replied: {\em Hypothesis non fingo!}  That is, ``I have no idea!``   We don`t go any deeper than Newton\textquotesingle s laws.} 

\begin{figure} 
  \centering
     \includegraphics[width=70mm]{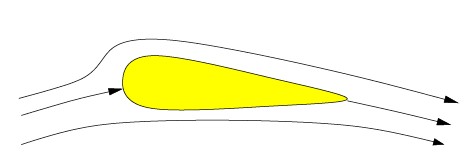}
     \caption{Bending of the airflow by an airfoil.}
     \label{fig:airfoil_flow}
\end{figure}

Figure \ref{fig:airfoil_flow} is a drawing of an airfoil that causes bending of the flow.  Notice that the bending is not just the deflection of the air by the lower surface of the airfoil.  The flow along the top is bent also.  The cause of the bending of the flow over the top of the wing is also the cause of the Coand\v{a} effect.  The behavior of the flow far from the surface of the airfoil is affected by the complex interaction of the surface of the airfoil with the molecules making up the flow.

The pressure at the surface of the airfoil, not the third law behavior of the flow far from the surface, is what actually causes the lift and drag.  If a method could be developed to compute this pressure, then lift an drag could be computed from first principles.

\subsection{Using Newton\textquotesingle s Second Law: Effects on the airfoil caused directly by air pressure}
The lift force is due to the differential in pressure between the bottom and the top of the airfoil.  If we assume that the air is approximately an ideal gas, Equations \ref{eq:IDgas} and \ref{eq:density}  show that the pressure, $p,$ at a given temperature and the mass density, $\rho,$ are proportional.  

First, assuming an ideal gas, notice that  the mass density of the air is

$$\rho = {{nNm} \over V},$$

where

\begin{itemize}
\renewcommand{\labelitemi}{$\ $}
\item{$n$ is the number of moles of the gas in volume $V$,}
\item{$N \sim 6.02 \times 10^{23}$ is Avogadro\textquotesingle s number,}
\item{$m$ is the mass of one particle.}
\end{itemize}

This leads to,

\begin{equation}
\label{eq:p_rho}
p = {\rho \over {Nm}} \times RT
\end{equation}

where

\begin{itemize}
\renewcommand{\labelitemi}{$\ $}
\item{$p$ is the pressure,}
\item{$R$ is the gas constant and}
\item{$T$ is the Kelvin temperature.}
\end{itemize}

Some standard methods of  increasing the pressure on the bottom of the wing are angle of attack, of course, and trailing edge flaps.

But we have seen above that the most important contribution to lift, in the usual range of angles of attack, is the suction at the top of the wing.  This decrease in pressure as the air density is reduced is the cause of the Coand\v{a} effect there.  It is due to the decrease in $\rho,$ the decrease in the particle density in the boundary layer.

In a given constant volume, $V,$ just above the surface of the wing, particles enter and leave.  Let $\eta(s)$ be the particle density at $s$ on the top of the wing.  According to Equation \ref{eq:IDgas} we can write

\begin{equation}
\label{eq:p_eta}
p = \eta kT,
\end{equation}

where $\eta$ is the particle density.   We submit the following model for the particle density in the boundary layer of a surface curving away from the main flow,

\begin{equation}
\label{eq:eta_model}
{\eta (s)} = \eta_0 \Im \left ({{v_0} \over v(s)} \right ) \times \Pi(s).
\end{equation}

where

\begin{itemize}
\renewcommand{\labelitemi}{$\ $}
\item{$\eta (s)$ is the particle density at point $s$ on the surface,}
\item{$\eta_0$ is the ambient particle density,}
\item{$\Im$ is a dimensionless function,}
\item{$v_0$ is the velocity of the main flow,}
\item{$v(s)$ is the velocity of the flow just outside the boundary layer ($bl$) at point $s$ on the surface and}
\item{$\Pi(s) = Pr(scatter~into~bl)$ is the probability that a particle in the boundary layer will be scattered back into it.}
\end{itemize}

\begin{figure} 
  \centering
     \includegraphics[width=70mm]{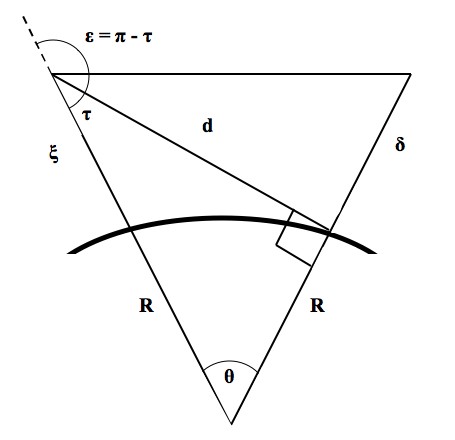}
     \caption{Coand\v{a} effect geometry.}
     \label{fig:Coanda_geometry}
\end{figure}

Figure \ref{fig:Coanda_geometry} depicts a portion of the curved part of the airfoil surface.

The main flow is from the left.  Since they are stagnant or nearly so, the particles in the boundary layer, thickness $\delta,$ will be scattered.  We assume that all scattering is in the direction of the main flow and all angles are equally probable.   Those particles that are scattered into angle $\tau$ remain in the boundary layer.  The probability for a boundary layer particle to be scattered back into the boundary layer is
 
\begin{equation}
\label{eq:Pi_model}
\Pi (s) ={1 \over \pi} \int_R^{R + \delta}  \tau (\xi, R(s)) d\xi .
\end{equation}

 Figure \ref{fig:Coanda_geometry} yields, 
  
\begin{equation}
\label{eq:tau}
\tau (\xi , R(s)) = sin^{-1} \left [ R \over {R + \xi} \right ].
\end{equation}

Combining Equations \ref{eq:p_eta} through \ref{eq:tau} there results

\begin{equation}
\label{eq:suction}
p(s) = {{{\eta_0}kT} \over \pi} \times \Im \left ( {v_0} \over v(s) \right ) \int_R^{R + \delta} sin^{-1} \left [ R(s) \over {R + \xi} \right ] d\xi
\end{equation}

for the pressure on the top of the curved wing.  At or near zero angle of attack  the pressure on the bottom of a flat wing is approximately ambient so the total lift force acting on the wing is

\begin{equation}
\label{eq:direct_lift}
{\mathbf F}_{direct}  = {{{\eta_0}kT} \over \pi} \left \{ {\mathbf S} - \int_{\uparrow} \left [ \Im \left ( {v_0} \over v(s) \right ) \int_R^{R + \delta} sin^{-1} \left [ R(s) \over {R + \xi} \right ] d\xi \right ] d{\mathbf s} \right \}.
\end{equation}

where $\uparrow$ indicates the top surface of the wing and ${\mathbf S}$ is the area per unit span of  the bottom surface.  Since we consider the simple case of a flat bottom surface parallel to the flow, the force there is just the ambient pressure times the area.  The pressure on the upper surface, lowered by the action of the main flow shearing past the curved surface, is responsible for most of the lift at zero angle of attack, $\Phi = 0$.

Figure \ref{fig:Coanda-flow} in Section \ref{Flfl} hints that the flow, $v(s),$ over the curved surface consists mainly of boundary layer particles activated by interaction with the main flow.  The function $\Im$ needs to be determined.  Its form will have to do with the detailed structure of the surface and the shape of the molecules of the flow.

\section{Stalling wing}
As the flow velocity increases, the secondary collisions that affect the flow particles slowed by collisions with boundary layer particles cause the slowed flow particles to be more violently knocked back toward the surface by the main flow.  When they start to hit the surface itself, the forces on the surface there increase (see panel 3 in Figure \ref{fig:WingFlow}).  The wing is stalling.  As the velocity increases still further, the flow near\footnote{At the surface the flow velocity is much less than the main flow velocity.  The derivative of the flow velocity in the direction normal to the surface vanishes at the so-called separation point on the surface.  Downstream of this point the particles just above the surface are flowing in reverse of the flow.} the surface reverses itself and flows back along the wing and also increases the boundary layer population, and hence the pressure, there.  The flow is said to separate at the point where the backward flow rate equals the forward fluid velocity.  Downstream from this point, a vortex has formed. 

\chapter{Heat Engines: 
Mechanisms that create order from chaos}
As opposed to an orderly collective unidirectional movement, which does work, heat is the {\em chaotic} motion of molecules.  Mechanisms that convert heat energy into the mechanical energy of collective motion are called heat engines.  Examples are Hero\textquotesingle s s engine (the ancient version of a jet or rocket engine), steam engines and internal combustion engines.   

\begin{figure}[!htb] 
  \centering
      \includegraphics[width=70mm]{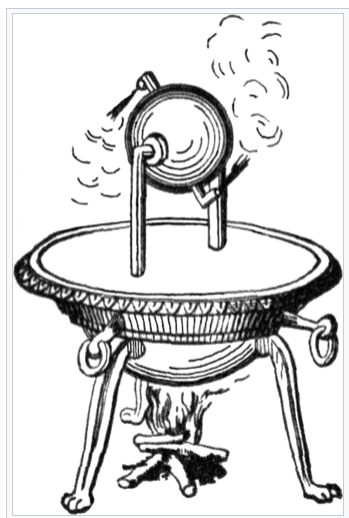}
      \caption{Hero\textquotesingle s s engine: the Aeolipile}
      \label{fig:aeolipile}
\end{figure}

Consider gas in an insulated container and that the container is divided into $N$  cells with $n^i$  the number of molecules in the $i^{th}$ cell. So that the total number, $\Pi$, of particles is 

$$\Pi =  \sum^{N}_jn^{j}$$

Now add an amount of  kinetic energy, $\Delta Q$, where,

$$\Delta Q = \sum_i^N \Delta Q^{i} $$

and  is $\Delta Q^i$ the energy added to the $i^{th}$  cell. 

$$\Delta Q^i = {1 \over 2} m \sum_{j}^{n^i} { {\mathbf v}}_{j}^i \cdot { {\mathbf v}}_{j}^i$$.

The mass of each particle is $m$ and  ${\mathbf v}_{j}^{i}$ is the velocity vector of the $j^{th}$ molecule in the $i^{th}$ cell. 

The total work, $W$, done by the gas is, 

$$W= \sum^N_i W^i$$

where,

$$W^i = {1 \over 2} m \big[ \sum_{j}^{n^i} { {\mathbf v}}_{j}^i\big] ^2$$

is the work done by the gas in the $i^{th}$ cell. Note that this is a vector sum. If the vector sum of the particle velocities in a given cell is zero, this cell does no work and all the kinetic energy there is heat energy.  

This equation reflects the fact that work is a collective activity of the molecules. The gas in the  $i^{th}$ cell can only do work if the vector sum of its particles\textquotesingle  velocities is non-zero, i.e. that the particles are primarily moving in the same direction.  

For each cell the first law of Thermodynamics gives the change in energy, $\Delta U^i$, in the $i^{th}$  cell as:

$$\Delta U^i = {(P.E. + K.E.)}^i = \Delta {\Xi}^i + \Delta  Q^i - W^i $$
$$\Delta U^i= \Delta {\Xi}^i + {1 \over 2} m \big[ \sum^{n^i}_j ({\mathbf v}^i_j)^2 - \big( \sum^{n^i}_j {\mathbf v}^i_j \big)^2 \big]$$

$\Delta \Xi^i$  is the change in potential energy of the $i^{th}$ cell due to change an external field, such as gravity, $\Delta Q^i$ is the heat energy added to the $i^{th}$ cell and $W^i$ is the work done by the gas in the  cell (hence the minus sign) as a result of any non-zero vector sum of the molecule velocities in the  cell.  is energy that leaves the system in the form of mechanical work. 

\begin{center}*********************** \end{center}

The purpose of a heat engine is to convert heat energy into energy in another form.  The most common form is mechanical energy, 

$$\Delta W = {\mathbf F} \cdot \Delta {\mathbf s} = P \Delta 
V$$ 

Let\textquotesingle s s think now about an isothermal process in a cell of hot gas which is fitted with a moveable wall, a piston. The piston is usually connected to a mechanical linkage which converts heat energy to mechanical work. When work is done by the gas on the piston, the total energy is conserved but it is apportioned between heat and work so now, assuming that there is no change in potential energy, i.e., $\Xi = 0$,

$$\Delta U = \Delta Q - W = \sum_i^N  \Delta Q^i -  \int_{piston} {\mathbf F} \cdot d{\mathbf s}$$

As the work, $W$, occurs, the energy, $U$, decreases. At any given instant, though, the piston is affected only by those molecules that actually collide with it.  So that momentum is conserved, the  $x$-components of their velocities are reversed and decreased as the piston moves. Heat energy is converted into the mechanical energy of the moving piston and the gas cools, but just at the surface of the piston. The cooling of the entire volume is by conduction as the more energetic molecules away from the piston encounter the ones cooled as work is done on the piston and by convection as the cooler molecules are swept away from the piston and into the volume by turbulent flow.  

The force on the piston is due to the molecules that strike it, i.e. those adjacent the piston which have positive $x$-components of velocity.

\begin{center} *********************** \end{center}

Another type of heat engine the so-called reaction engine. Hero\textquotesingle s s engine (Figure \ref{fig:aeolipile}) as well as rocket and jet engines, work as a result of the conversion of the random heat motion in a gas to a more orderly motion of a directed exhaust, which does work.  

The detailed behavior of a rocket motor is not as obvious as one might think. The thrust is due to the difference in pressure between that at the surface of the motor\textquotesingle s s chamber opposite the plane of the  exhaust nozzle and that at the exhaust throat.  The pressure field in the motor, then, is not uniform and, depending on the detailed geometry of the motor\textquotesingle s s chamber, may have a complicated position dependence due to non-uniformity of the temperature field there.  

The ideal gas law for a gas in equilibrium at pressure $p$ and with volume  $V$ , as was shown by Rudolf Clausius in 1857, is

$$pV = {1 \over 3}Nmv^2$$

where  here $V$ denotes the volume of the gas, $N$ is the total number of molecules in the system, $m$ is the mass of a molecule and $v^2 = {\mathbf v} \cdot {\mathbf v}$ is the square of the velocity of a molecule. If one makes the approximation that the ideal gas law obtains within a rocket motor and that there is one kind of molecule, the pressure there, described by:

$$p(\mathbf x) = {\rho(\mathbf x)  \over m} kT(\mathbf x) = {1 \over 3} \rho(\mathbf x) \big[
{\mathbf v}({\mathbf x}) \cdot {\mathbf v}(\mathbf x)\big]$$

where $\rho ({\mathbf x})$ is the particle density and $k$  is Boltzmann\textquotesingle s s constant.  

The thrust of a motor without a shaped nozzle, then, is:

$${\mathbf F} = p_{back~wall} \times {\mathbf A}_{back~wall} - p_{exit~orifice} \times {\mathbf A}_{exit~orifice}$$.  

\begin{figure} 
  \centering
     \includegraphics[width=70mm]{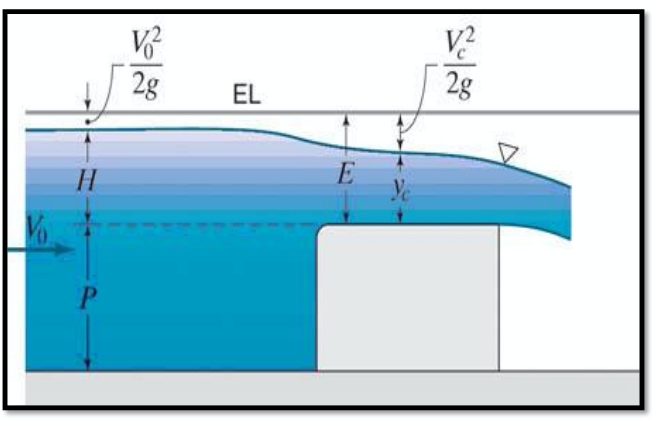}
     \caption{Broad-Crested Weir}
     \label{fig:Weir}
\end{figure}

\begin{center} ************************* \end{center} 

A system mathematically analogous to the reaction motor is the anti-siphon standpipe weir. In the past, ground flow irrigation systems used anti-siphon devices called standpipe weirs. The water delivered to the field or orchard entered the weir from underground and flowed up and over a spillway open to the atmosphere. The spillway then lead back underground to the pipes that delivered the water to the crops. The equation describing the adiabatic\footnote{ Adiabatic means that there is no heat input or loss, i.e. $\Delta Q = 0$.} flow of water in such a weir is:

$$U= \Xi - W = \rho g  (H - y(x)) - {1 \over 2} \rho V^2(x,y) = 0$$

or

$$y(x) = H - {{V^2(x,y)} \over {2g}}$$

 ($V^2(y)$ is the square of the fluid\textquotesingle s collective velocity at height $y$.) Notice that this equation is the same as First Law equation except that the potential energy, $\Xi$, in the gravitational field is non-zero and $\Delta Q = 0$. Under the approximation that the water is incompressible, i.e. that $\rho$ is constant, the pressure difference due to the water in the region $H$ will result in a pressure on the rear wall of the weir. 

$$p_i({\mathbf x})
= {1 \over 3} \rho \big[ {\mathbf v}_i({\mathbf x}) \cdot {\mathbf v}_i({\mathbf x})\big]$$

Here, $p_i({\mathbf x})$ and ${\mathbf v}_i({\mathbf x})$ are the pressure and the velocity of particles in the $i^{th}$ coarse-grain cell, the cell at  near the rear wall of the bottom-fed weir. It is assumed that the magnitudes of the velocities of all the particles are equal and that their directions are random,  i.e. the vector sum of the velocities there is zero, as is the assumption for Clausius\textquotesingle \,equation. At the rear wall there is approximately zero net flow, $V$ .  There, the pressure is maximum. On the other hand, the water level drops at the spillway where the vector average of the velocities in the cells is not zero. Work is done on the water  by the gravitational field and work can be done by the water as its gravitational potential energy is converted into a collective and orderly kinetic energy of the flow of the molecules.  The average molecular velocity will be in the direction of the exiting flow. The pressure there is minimum.  The difference between the rear wall pressure and the pressure at the spillway results in a net force on the rear wall. The shape of the water surface varies as the flow approaches the spillway.  

The above equation also holds for a rocket motor.  The pressure on the back wall is maximum and the pressure in the motor decreases to some minimum as the exit orifice is approached. The exact ``shape`` of the 3-dimensional pressure field in the rocket motor depends on the shape and orientation of the exit orifice and the details of the inner shape of the motor which, for a solid propellant motor changes during the ``burn,``  just as the contour of the surface of the water flowing from a weir standpipe depends on the exact shape of the exit channel.

\chapter{Devices utilizing the attributes of flow at the particle level}
As will be seen below, engineers routinely exploit the particle behavior of fluids even though there is no way to calculate these effects just on the basis of fluid as a primitive substance.  An example is the design of high-performance internal combustion engines.  Far from making assumptions of equilibrium and steady flow, these engineers have learned how to exploit the nonlinear and turbulent behavior of real fluids, e.g., the swirling flow of gasses both in the combustion chamber and the exhaust system. These engines, including intake and exhaust systems, are actually highly tuned, integrated high-Q resonant systems, that is, their power output is highly dependent on the reciprocating frequency, the RPM.  The shape of the piston dome, the valve arrangement and the tuned exhaust are designed to work as a unit.  The effects are quite spectacular.
\section{Rocket engine diffuser} \label{REdiffuser}
A rocket engine is composed of three basic parts: the pressure chamber, the throat and the exit horn or diffuser.  The following is Bernoulli\textquotesingle s s equation applied to the thrust of a rocket engine with no diffuser, just an exit orifice of area $A$:

\begin{equation}
\label{eq:pengine}
p _{engine} = p _{orifice} + {1 \over 2} \rho{v_{exhaust}}^2
\end{equation}

The thrust, then, is

\begin{equation}
\label{eq:thrust}
Thrust = (p _{engine} - p _{orifice}) \times A = A\times({1 \over 2} \rho{v_{exhaust}}^2).
\end{equation}

The thrust is caused by the pressure, i.e., particle collisions on the side of the engine away from the throat, that is not offset by the throat pressure.  It is not {\em caused} by the mass flow rate in the exhaust. If it were confined with no exit orifice, the gas would have provided pressure offsetting that at the other end of the engine and the thrust would be zero in the above equation.    In the absence of a diffuser, thanks to Bernoulli\textquotesingle s s equation, the thrust can be calculated using the exhaust velocity and density but these are not what {\em cause} the thrust.  

The thrust of a rocket engine without diffuser is easy to calculate but is much enhanced by the non-Bernoulli flow in the diffuser. 

\subsection{The diffuser}
\begin{figure} 
  \centering
     \includegraphics[width=70mm]{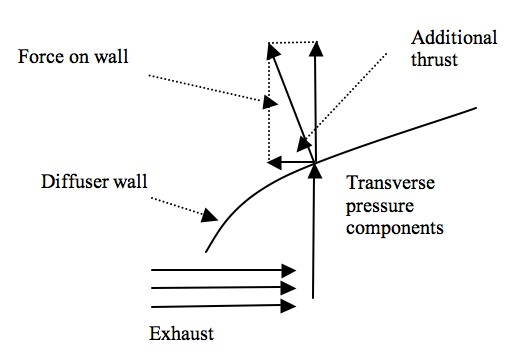}
     \caption{Detail at the Diffuser Wall}
     \label{fig:diffuser_detail}
\end{figure}

The thrust of a rocket engine is substantially increased if the exhaust gases exit into a  diffuser\cite{schlichting}.   Diffusers are prominent in films of rocket launches and may be examined in aeronautical museums like Le Bourget outside Paris, France.   As the rocket exhaust exits into a parabolic chamber, it spreads to fill the entire volume of the chamber\cite{schlichting} and, due to the fact that the gases in the diffuser, i.e., downstream from the orifice, are highly interacting, Bernoulli\textquotesingle s s equation no longer holds.  A symptom of this breakdown of Bernoulli flow is that the exhaust velocity is no longer uniform across the surfaces normal to the exhaust flow.  The exhaust spreads because of the large transverse thermal velocity components of the hot gases.   This can be seen as a sort of transverse pressure as is measured in a Venturi tube.  Combining Equations \ref{eq:pvessel} and \ref{eq:pengine} we see that just at the orifice 

$$p _{orifice} =  {1 \over 2} \rho\sum (v_{\it x}^2 + v_{\it y}^2).$$  

where the sum is over all the particles at the orifice.  It is this transverse pressure that drives the exhaust to the diffuser wall.   As the hot gases collide with the diffuser wall, they produce a force normal to that surface.  The shape of the diffuser is designed to provide force components in the direction of the rocket\textquotesingle s s velocity.  (See Figure \ref{fig:diffuser_detail}.)

In the absence of the diffuser, the extremely energetic exhaust gases would carry much of their energy away.  In fact, one can see this as the rocket plume spreads beyond the diffuser at high altitude in a sort of umbrella shape.  The diffuser, then, extracts work from the gases before they exit beyond the rocket.  The analogous situation in an internal combustion engine is to delay the opening of the exhaust valves in order to extract more work from the hot gas in the combustion chamber.  

An interesting feature of some diffusers, like those of the French Ariane rocket,  is a sort of rifling on the diffuser wall.  This rifling causes the gases to swirl, thus increasing the path length for the exiting particles.  This keeps them in contact with the wall longer allowing more of their heat energy to be converted into thrust.

\section{The high-bypass turbofan jet engine}
By the 1950s the turbojet had largely replaced the piston engine driven propeller as the main means of aircraft propulsion.  The advance in the design of heat-seeking missiles, however, was becoming a serious threat to military aircraft due to the high temperature of the jet exhaust.  The thought occurred to engine designers that a sheath of cooler air would mask the heat signature of the jet exhaust so the heat-seeking missiles could could not ``see`` it.  This bypassing air, however, did not only shield the exhaust but it increased dramatically the overall efficiency of the jet engine in subsonic flight.  Thus was born a true breakthrough in jet engine design, a very useful spin-off of technology originally intended only for military use.

\begin{figure}[!htb] 
\centering
      \includegraphics[width=70mm] {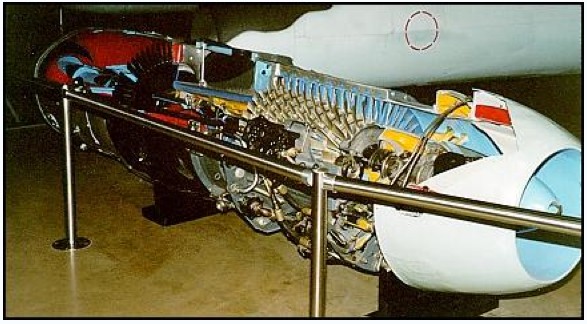} 
      \caption{Early turbojet}
      \label{fig:oldturbojet}
\end{figure}

Before the development of the modern high-bypass turbofan\footnote{In high-bypass turbofan engines, most of the air entering the intake cowl bypasses the the jet engine itself.  This air is compressed somewhat by the shape of the entry cowl and its attendant bow-wave.  The compressed air is propelled by a fan driven by a conventional turbojet engine and then exits the cowl in an annulus at the rear. An added benefit in using any ducted fan is that the vortices at the ends of the blades are inhibited by the duct wall} jet engine, the rim of the entrance to the engine cowl was rather sharp.  (See Figure \ref{fig:oldturbojet}.)

\begin{figure}[!htb] 
\centering
      \includegraphics[width=70mm] {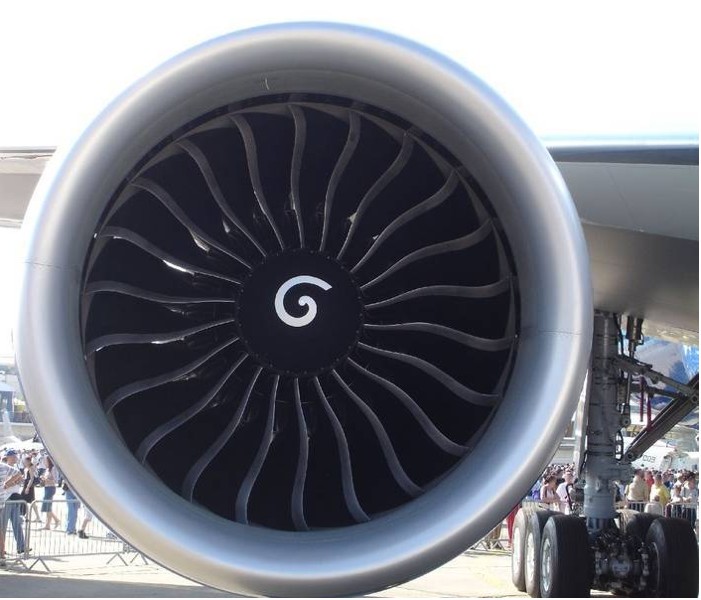} 
      \caption{The entrance cowl for an Airbus A380 turbofan engine}
      \label{fig:turbofan}
\end{figure}

\begin{figure}[!htb] 
\centering
      \includegraphics[width=70mm] {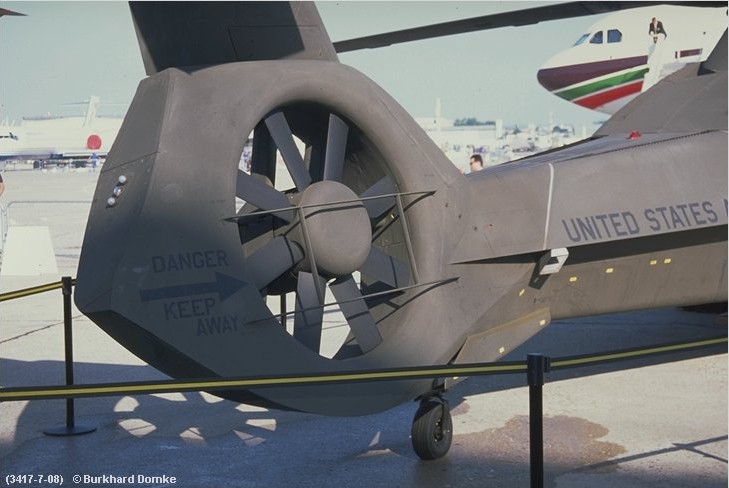} 
      \caption{Ducted fan tailrotor}
      \label{fig:Ducted_fan_tailrotor}
\end{figure}

Modern turbofans (Figure \ref{fig:turbofan}) have a gently curved entrance duct.  This may seem trivial but it has the effect of greatly increasing the entrance aperture for the turbofan due to the  Coand\v{a} effect.  This cowl creates a bow wave as it moves through the air.  The presence this bow wave at subsonic speeds\footnote{At supersonic speeds this effect disappears because the aircraft has outrun its bow wave.} creates a flow pattern that precedes the aircraft.  Examination of the third panel of Figure \ref{fig:WingFlow} shows how the presence of a curved surface in an air flow causes what might be called Coand\v{a} entrainment of air from outside. This entrainment compresses the air {\em in front} of the fan so that the fan is moving air of an increased density, $\rho.$  This increased density and the boost in velocity provided by the fan, according to Newton\textquotesingle s s third law, results in an increase in thrust pressure $p$.

A similar Coand\v{a} entrainment enhances the performance of the shrouded tail rotors (Figure \ref{fig:Ducted_fan_tailrotor}) that are used on some helicopters.\cite{shroud}  Figure \ref{fig:Ducted_fan_tailrotor} is used with the kind permission of Burkhard Domke.\footnote{http://www.b-domke.de/}

\section{The vortex refrigerator}\label{VortexRefrig}
A device called a vortex refrigerator consists of a cylindrical chamber into one end of which a gas is injected tangentially.  Gas is then drawn off, cold, from the axis of the cylinder and hot from its periphery.  Along the length of the cylinder the cold molecules are separated from the hot by the vortex process outlined in Figure \ref{fig:VortexProcess}.  

The temperature of the cold air exiting from a vortex generator can be substantially below that of the compressed air at the inlet.\footnote{See the websites of manufacturers of vortex refrigerators: \newline\hspace*{25 pt}http://www.exair.com and also http://www.airtxinternational.com }  Although the vortex process described above would serve as a sort of ``Maxwell\textquotesingle s s d{\ae}mon`` to separate the cold particles from the hot there may be some other process that changes the distribution of the particles\textquotesingle \, energies, e.g., increases its variance.  It is easy to imagine that some of the translational energy of the inlet air would be converted into heat of the exhausting air but it is more difficult to understand how the cold air exiting could be as much as $28^\circ - 50^\circ$ centigrade below the inlet temperature.  Unfortunately, the measured temperature is a macroscopic quantity.  It would be interesting to see if the inlet energy distribution, i.e. kinetic plus heat, is the sum of the cold and hot exhaust distributions.

\section{Hurricanes}\label{hurricanes}The centers of hurricanes are regions of low pressure.  In the great hurricane of 1900 that struck Galveston, Texas, the pressure was the lowest ever recorded up to that time, 936 millibars.\footnote{Average standard pressure is 1013 millibars.}  The pressure recorded in the eye of hurricane Katrina which hit the coast of the Gulf of Mexico in 2005 was even lower than this at 920 millibars.  If the hurricane is over water, this low pressure causes {\it storm surge.}  The water in the center of the hurricane is pushed up because of the low pressure there and the higher pressure outside the center.  In the case of Galveston and hurricane Katrina and most other hurricanes, this storm surge caused most of the damage to the cities.  The mechanism that causes the structure of hurricanes and cyclones is not fully understood\footnote{See the Louisiana Homeland Security website at: \newline\hspace*{25 pt}
http://www.ohsep.louisiana.gov/factsheets/FactsAboutHurricaneEye.htm } but it certainly is a result of complex particle interactions.  The flow is definitely not steady so Bernoulli\textquotesingle s s equation will not hold.

\section{The Dyson Air Multiplier\textsuperscript{\texttrademark} fan}\label{DysonCool}
\begin{figure}[!htb] 
\centering
      \includegraphics[width=70mm] {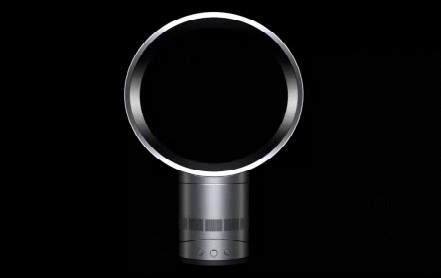} 
      \caption{The Dyson Air Multiplier\textsuperscript{\texttrademark} fan}
      \label{fig:DysonFan0}
\end{figure}

An interesting exploitation of the Coand\v{a} effect is the desk fan developed by James Dyson.\cite{Dyson}  In Figure \ref{fig:DysonFan0} perhaps the most striking aspect is the lack of visible fan blades.  Air is drawn into a plenum by an impeller in the base and is expelled at high speed from a slot around the large annulus.    

\begin{figure}[!htb] 
\centering
      \includegraphics[width=70mm] {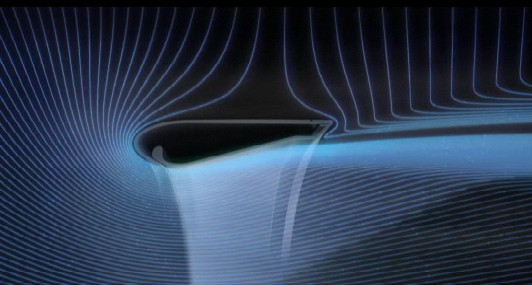} 
      \caption{Operation of the Dyson Air Multiplier\textsuperscript{\texttrademark} fan}
      \label{fig:DysonFan}
\end{figure}

 This air is directed over a curved surface shaped like an airfoil, creating a low pressure on the inside of the annulus as shown in Figure \ref{fig:forces_hepperle} of Section (\ref{lift_calc}).  Figure \ref{fig:DysonFan} shows the resultant Coand\v{a} flow.
 
\begin{figure}[!htb] 
\centering
      \includegraphics[width=70mm] {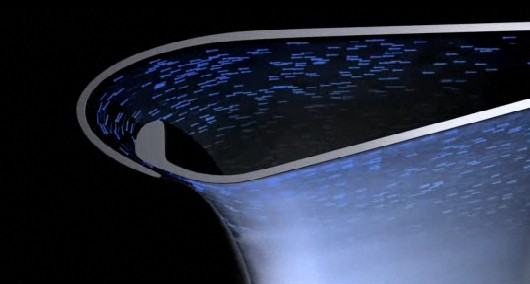} 
      \caption{Cross section of the Dyson Air Multiplier\textsuperscript{\texttrademark} fan}
      \label{fig:DysonFan1}
\end{figure}

Note that the volume rate of the air drawn into the annulus is much larger than the flow from the impeller in the base.  Figure \ref{fig:DysonFan1} shows a cross section of the annulus with detail of the annular slot.

There is another quite interesting effect. The Dyson Air Multiplier\textsuperscript{\texttrademark} (Section \ref{DysonCool}) produces is local cooling without refrigeration (See the discussion of the Venturi tube above.).

\section{Spinning objects in the flow: the Magnus effect}
\begin{figure}[!htb] 
\centering
      \includegraphics[width=70mm] {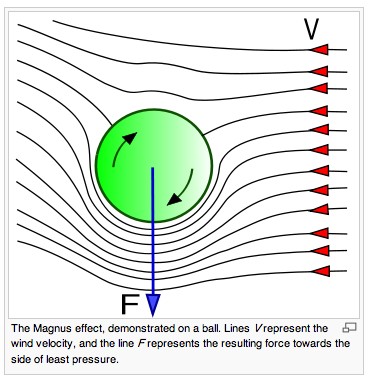} 
      \caption{The Magnus effect}
      \label{fig:Magnus}
\end{figure}

The Magnus effect is well-known by players of tennis, ping-pong, baseball, soccer and volleyball.  It is illustrated in Figure \ref{fig:Magnus}.   Discovered by Heinrich Gustav Magnus (1802 - 1870), the Magnus effect has been used to power a ship across the Atlantic.\footnote{See a picture of the ship that Anton Flettner built in the 1920s:  \\ http://www.tecsoc.org/pubs/history/2002/may9.htm } 

A rotating object in a flow will generate a differential pressure which will produce a force on the object normal to its spin axis.  The pressure is higher on the side rotating into the flow than on the side rotating with the flow.   The effect is caused as the boundary layer dragged along by the object is pressurized by the main flow.  This pressurization is caused by the collisions of the flow particles with the boundary layer particles and the surface structure of the object.  The furry surface of a new tennis ball enhances this effect.  Another means of pressurizing the boundary layer, used on some Formula One race cars, is the Gurney flap (See Section \ref{flaps} below).

\section{Gurney and Fowler flaps}\label{flaps}
\begin{figure}[!htb] 
\centering
      \includegraphics[width=90mm] {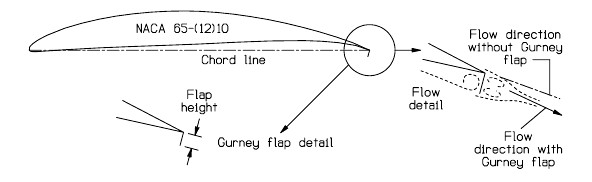} 
      \caption{Gurney flap}
      \label{fig:GurneyFlap}
\end{figure}

Although the Gurney flap was actually invented by Edward F. Zaparka in the 1930s\cite{gurney}, race driver and race car builder Dan Gurney accidentally rediscovered it in 1971.  The height of simplicity, it is a length of aluminum angle iron bolted to the trailing edge of an airfoil.  It causes an increased pressure on the side of the airfoil from which it projects.  It is easy to see that it functions as a dam, trapping air from the flow, thus increasing the pressure on that side of the airfoil.  For a race car this is done to increase the force down on the wheels, decreasing the chance for wheel spin and increasing traction for the rear wheels. Figure \ref{fig:GurneyFlap} details this simple but very effective device as used to increase lift on an airfoil.

Ordinary Fowler flaps\cite{flaps}, common on commercial airliners, cause a similar damming of the flow on takeoff and landing, where maximum lift is needed.  The top surface of a Fowler flap also utilizes the Coand\v{a} effect to enhance lift.  ``Slotted`` Fowler flaps direct some of the high-pressure air under the wing over the rear flap sections, thus enhancing the Coand\v{a} effect there. 

\section{Slots and slats}\label{slots}
A slot\cite{flaps} is a gap between a slat and the leading edge of a wing or between the sections of the Fowler flaps at the trailing edge.

\begin{figure}[!htb] 
\centering
      \includegraphics[width=90mm] {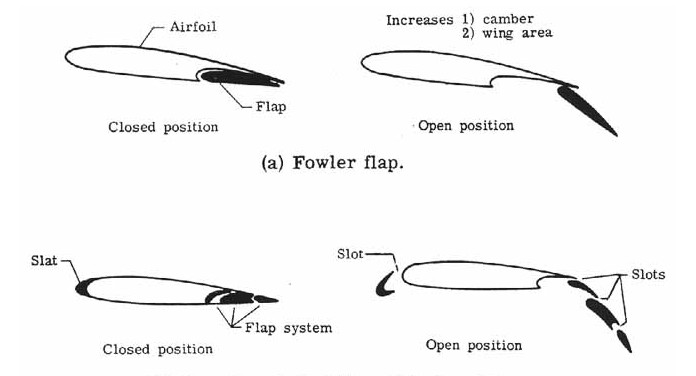} 
      \caption{High-lift wing devices}
      \label{fig:flaps}
\end{figure}

On takeoff and landing, some airliners, in addition to Fowler flaps, use leading edge slats and slots\cite{flaps}.  Designs differ from manufacturer to manufacturer but a leading edge mechanism of some kind is always incorporated into the design of the wings of airliners.  Leading edge slats and slots direct the upwash from the bottom of the wing at angle of attack over the leading edge of the airfoil.   

A simple experiment will demonstrate upwash.  Hold a flat plate at a $10^\circ$ or $20^\circ$ angle with respect to a smooth stream of water.  As the water strikes it, it climbs up the plate against the direction of the water flow.  This effect is due to the interaction of the water molecules with the microscopic protuberances on the surface of the plate.  Another way of stating this is to say that the effect is due to the viscosity of the water.  

As the slat descends into the air flowing beneath the wing, it opens a slot. (See the 4th panel of Figure (\ref{fig:flaps}.)  This combination directs the upwash air smoothly onto the top surface of the wing causing a Coand\v{a} effect there which increases lift.  Instead of slots and slats, some Airbus planes use vortex spoilers on the leading edges of their wings.  These spoilers, like slats, are deployed only on takeoff and landing and serve to prevent a stalling vortex from forming along the leading edge of the wing.

\section{Slurries}

A slurry is a suspension of insoluble particles.  Normally one thinks of the particles as suspended in water as in a mudslide or a cement slurry that can be pumped as if it were a thick liquid.  Slurries transport the sand in river bottoms and create the sandbars that accrue in the rivers.   

Quicksand is what could be called a bistable slurry (my term).  It is a mixture of water and fine sand or clay particles.  Undisturbed, quicksand appears as though it were solid ground (state \#1).  In this stage the slurry particles rest on one another due to gravity and, since the particles are denser than water, they tend to settle below the surface of the water. Even so, at every level there are water molecules surrounding the heavier particles. When it is disturbed, e.g. by an earthquake or by someone\textquotesingle s s foot, the quicksand rapidly transforms into a liquid suspension (state \#2).  In a sense what happens is a phase change, like the melting of ice. The effective viscosity of the slurry reduces radically and quickly. This takes place as the disturbance spreads through the stage \#1 mixture at the speed of sound there. This speed can be over 1400 m/s, depending on the temperature and the shape, size and mass of the slurry particles. The sound wave, a compression-rarefaction longitudinal wave, spreads through the quicksand, forcing the slurry particles apart as the water molecules move between them.  
  
These water liquid slurries are quite familiar but there are other kinds of slurries.  

Smoke is a slurry of particles entrained in air.  As in the case of the river slurries, the soot particles deposit onto the surfaces with which the smoke comes in contact.  For example, in Paris the towers, churches and other monuments must be cleaned periodically because of the deposits of smoke particles, in large part due to diesel exhaust from cars.  When the entraining air is still hot, the smoke is buoyed by the air and as the hot air rises, so do the smoke particles.  A very interesting kind of gas slurry is the superheated cloud that accompanies a volcanic eruption.  

The particles in the pyroclastic cloud that accompanies the violent eruption of a volcano, buoyed and swept along by the hot gases, are created as the bubbles of lava magma burst.  In a private communication from Dr. Haraldur Sigurdsson of the University of Rhode Island, he describes the pyroclastic phenomenon as it occurred in the eruption of mount Vesuvius in 79AD:

{\em \quotation

The temperature of the erupting magma and gases [was] around 900 $^ \circ C$ for Vesuvius. The gas-rich magma fragments on the way to the surface, due to bubble expansion and bubble bursting about 5400 metres below the surface in the volcanic vent. The resulting mixture of fragmented magma (volcanic ash) and gases has a relatively low bulk density, less than that of air initially. Upon rising above the volcano the mixture further mixes with or draws in atmospheric air and a hot-air balloon effect occurs, with the entire eruption column rising to levels where the eruption cloud is neutrally buoyant. 

If mixing with air is inefficient, such as in the case when the eruption rate [is high] (of the order of [a] billion kg/second) then the eruption column is always more dense than air and collapses and flows as a pyroclastic flow along the ground, hot, fast and deadly. That was the case at Vesuvius.

\endquotation}

An interesting aspect of pyroclastic clouds and pyroclastic flow is mentioned in Dr. Sigurdsson\textquotesingle s s last paragraph.  When the mixing with the air and hot gasses is inefficient, then, rather than rising like a smoke column, the slurry is too dense to be buoyed by the atmosphere and the cloud flows along the ground.  This cloud, in pyroclastic flow, is still extremely hot, $\sim 900 ^ \circ C$.  As it flows over water it will cause the water to boil, or cause an automobile to burn, along with its contents, as shown in Figure \ref{fig:bus} below.  

\begin{figure}[!htb] 
\centering
      \includegraphics[width=90mm] {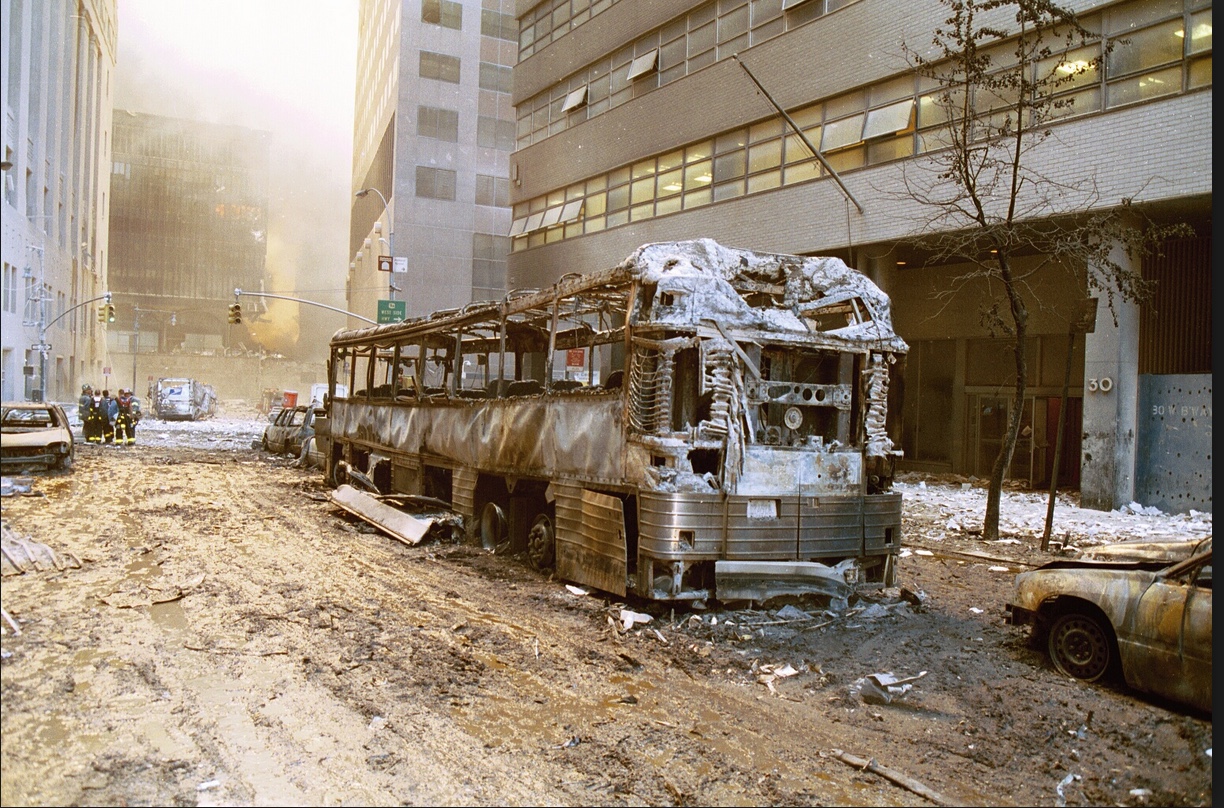} 
      \caption{Bus burned by the pyroclastic flow from the World Trade Center collapse on September 11, 2001.}
      \label{fig:bus}
\end{figure}

This happened in the pyroclastic flow that developed in New York on September 11, 2001 as the twin towers collapsed.  

When a volcano erupts violently a slurry is formed consisting of fine magma particles suspended in the hot gases.  This suspension develops as the Brownian motion of the gas is transmitted to the magma particles.  In addition to their kinetic energy the particles themselves have been heated internally by conduction and by radiation in the volcano.  If the number density of these particles in the pyroclastic flow is high enough, the flow becomes more dense than the surrounding air and so it hugs the surface of the ground or water, or whatever surface is low in the gravitational field.  It is the temperature of the particles themselves as well as the kinetic energy of the slurry particles that comprises total energy density of the pyroclastic flow. In the case of the collapse of the towers on September 11, 2001, the primary repository for the energy in the flow was the iron and steel droplets in the slurry, not the heat of the gas.  

\section{Summary}
This paper begins an investigation into how a fluid consisting of particles interacts with itself and with solid surfaces.   Although the mathematics of fluid dynamics is beautiful, it fails to explain many mysteries that appear in the behavior of common subsonic flows.  Many Physics books, for example, attempt to explain the phenomenon of subsonic lift by using Bernoulli\textquotesingle s s equation.  It is hoped that the brief treatment above will disabuse the reader from such an explanation. 

\section{Conclusion}
Anyone who has contemplated ocean waves in their magnificence or watched solitons marching upstream in channel flow cannot but be amazed at the beauty exhibited by the behavior of fluids, this in spite of difficulties that defy our ability to calculate their detailed behavior.  It is said that the first stage of understanding a phenomenon is that of its careful observation.  Perhaps the next is to attempt to formulate the behavior of the constituents of the phenomenon.   Prior to the turn of the 19$^{\textrm{th}}$ century, it was believed that fluids were fundamental entities, i.e., not composed of anything.  Since Einstein\textquotesingle s s 1905 paper on Brownian motion, however, it has been clear that fluids are composed of molecules.  To assume, then, at least for the sake of investigation, that all fluid behavior is caused by the interactions of these molecules and those of solid surfaces immersed in the fluid, seems natural.  These assumptions are useful even in the absence of a tractable mathematics to describe behavior at this level.  The current mathematical approaches, including the Navier-Stokes equations, make the fluid approximation\cite{landau} fundamental.  Some of the most baffling behavior of fluids, however, takes place in regimes where the fluid approximation is not valid.  

Perhaps, as computers become faster and with more and more memory, some of these behaviors will succumb to calculation.  In the meantime we observe and contemplate and we are amazed.

\newpage
\chapter{Appendices}
\section{Kinetic Theory recap}\label{KT recap}
Around 1827 Scottish botanist Robert Brown noticed with amazement that small pollen grains suspended in water moved ceaselessly and chaotically. At first he thought that the movement was due to the so-called''life force`` Later he and others noticed that grains of powdered stone also exhibited this motion so the mystery grew deeper.

It wasn`t until 1905 that Albert Einstein in one of his four seminal articles published that year, derived equations describing Brownian motion of macroscopically visible particles in terms of the statistics of the momentum transferred to these grains by much smaller, sub-microscopic particles in chaotic motion. This derivation brought life to the notion of molecules as physical entities instead of just convenient fictions.  

Before that fluids; water, oil, air, etc., were thought to be primitive objects, not made of anything more fundamental.  The energetic behavior of fluids was described by thermodynamics and the equations of fluid dynamics were used to describe their dynamical behavior. Even today modern fluid dynamics books begin with what is called the ''fluid approximation.`` They admit, of course, that the fluids are really made up of atoms, molecules, and in the case of a plasma, nuclei and electrons too. 

In order that the Navier-Stokes equations, already derived for fluids as fundamental entities, apply to this newly discovered reality, the fluid approximation is still made.  The Navier-Stokes equations were derived in the years between 1827 and 1845, prior to the results of Einstein. Nowhere in these equations is there a reference to particle behavior.

Although the notion that all matter is made up of atoms dates from Lucretius in 50 BCE, at the beginning of the twentieth century atoms were considered by scientific community to be purely hypothetical constructs, rather than real objects. An important task of the Kinetic Theory of Gases was to derive the properties of an ideal gas, assuming that a gas is a collection of molecules or atoms.  An ideal gas, i.e. a gas consisting of perfectly elastic spheres with no internal degrees of freedom, is composed of molecules that only affect one another when very close (colliding). The already well-known equation of state is

\[ pV = NkT = nRT \tag{1} \label{eq:eqst} \]

or

\[ p = {\rho \over m}kT \]

where:

\begin{itemize}
\renewcommand{\labelitemi}{$\ $}
\item{$p$ is the pressure}
\item{$V$ is the total volume of the gas}
\item{$N$is the number of molecules}
\item{$k$ is Boltzmann?` constant}
\item{$n$ is the number of moles of the gas}
\item{$R$ is the universal gas constant}
\item{$\rho$ is the density of particles}
\item{$m$ is the mass of one molecule}
\item{$T$ is the temperature in degrees K}
\end{itemize}

\begin{figure}[!htb] 
\centering
      \includegraphics[width=90mm] {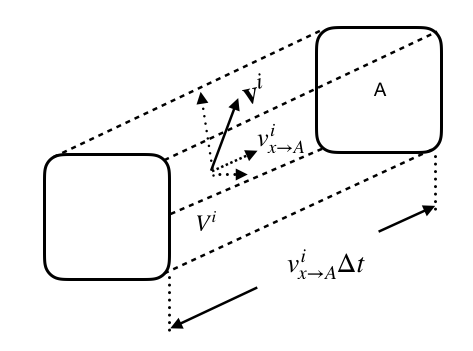} 
      \caption{Kinetic Pressure}
      \label{fig:kineticpressure}
\end{figure}

 \subsection{Derivation of the Ideal Gas Law}
 
 In the diagram, ${\mathbf v}^i$, is a velocity vector of one of the molecules of an ideal gas that happens to have the value of its $x$ position and its $x$-component $v^i_{x \rightarrow A}$ such that it will strike  $A$ in time $\Delta t$. It is counted in the volume $V^i$. $A$ is the area that the molecules in $V = \sum_i V^i$ with velocity components toward  will strike in time $\Delta t$. The sum above makes sense only because of the fact that the gas is made up of particles. The $\i^{th}$ volume is the sum of the volumes of all its particles. By the same token the volumes themselves are additive. This additivity, in turn, only makes sense because its measure is the molecular volume. This measure is called Lebesgue measure after the French mathematical Henri Lebesgue who first defined it. 
 
 Think of a pool table with all the balls randomly sitting on it. Using this measure, the total volume of, say, the even numbered balls is just the sum of the individual volumes swept out as the molecules move. Even though all the balls are on the table they can be broken up into separate and additive sets. The volumes of these sets are additive as long as no ball is counted more than once and the balls themselves don`t penetrate one another. This then means that the sets do not overlap. In the fluid approximation there would be only one volume, namely the product of the table\textquotesingle s area and the diameter of a ball. 

In order to calculate the force on the area $A$ in the diagram, it is necessary to find the number of particles that strike $A$ during the time interval $\Delta t$. These particles will all have velocity components in the $x$-direction but these components will be different for different particles, depending on how far they are from $A$. We will coarse-grain them into groups each group with a different mean value of its $x$-component of velocity.  Some of these will exit the volume laterally before striking $A$ but it is assumed that the gas is in equilibrium, in which case as many enter that way, on average, as leave.  

We begin by writing the total number of particles in the $i^{th}$ small volume, 

\[ V^i = A  v^i_{x \rightarrow A} \Delta t \],

where $\Delta t \leq \delta / \overline v$, and $\delta$  is the mean free path of the molecules between collisions and  $\overline v$ is the mean speed of the molecules. 

Call the number of molecules in this small volume $N^i$. Then,

\[ N^i = n^i  (A  v^i_{x \rightarrow A}) \Delta t = n^i V^i \],

where $n^i$ is the particle density for the $i^{th}$ group.

Under perfectly elastic collisions this group will contribute $2N^i mv^i_{x \rightarrow A}$
  to the momentum transferred to $A$.  The total momentum transferred during $\Delta t$, then, will be 

\[ \Delta \mu= 2 m \sum_i^{\# \ of \ groups} n^i  (A  v^i_{x \rightarrow A}) v^i_{x \rightarrow A} \Delta  t \].

For each group, however, only half of the particles with $x$-components of momentum are traveling towards $A$. According to the Equipartition Theorem, the average magnitude of the velocity vectors of the particles is independent of their directions, i.e. on average $v^i_{x \rightarrow A} = v^i_x  $, so 

\[ \Delta \mu = 2m \sum_i^{\# \ of \ groups} {1 \over 2} n^i ( A v^i_x)  (v^i_x  ) \Delta t ,\]

or

\[ \Delta \mu = m \sum_i^{\# \ of \ groups}   n^i A (v^i_x)^2 \Delta t = mA \sum_i^{\# \ of \ groups} {N^i \over V} (v^i_x)^2 \Delta t \]

But further, this shows that the pressure, $p$, the force, per unit area, i.e., 

\[ p = {F \over A} = {\Delta \mu \over A\Delta t} \]

is

\[ p = m \sum_i^{\# \ of \ groups} {N^i \over V} (v^i_x)^2 \]

Since in general,

\[ {\overline r} = {{\sum_i N^i r^i} \over N} \]

where

\[ N = \sum_i N^i ,\]

we can write

\[ {\overline {(v_x)^2}} = {\sum_i N^i (v^i_x)^2 \over N} . \]

So, in terms of the averages of the momentum transfers:

\[ p ={Nm \over V}\overline{{(v_x})^2} \]

By Pythagoras` Theorem

\[ {\mathbf v}^2 = v_x^2 + v_y^2 + v_z^2 ,\]

and since the gas is assumed to be in equilibrium, which is to say that there are negligible currents in the gas and also no temperature gradients, the Equipartition Theorem states that the terms on the right above are equal
so that, for example,

\[ {\mathbf v}^2 = 3v_x^2. \]

We can then write

\[ p = {1 \over 3} {Nm \over V}{\overline {({\mathbf v})^2}} = {1 \over 3} \rho {\overline {({\mathbf v})^2}}, \]

where $\rho = {Nm \over V}$ is the particle density. In any case we can write,

\[ p = {2 \over 3} {N \over V} \Big[ {1 \over 2} m {\overline {({\mathbf v})^2}} \Big]. \tag{2} \label{eq:ktpressure} ,\]

This explicitly shows that the pressure is proportional to the individual molecular energy density, 

\[ {1 \over 2} \rho {\overline {({\mathbf v})^2}}. \]

Before the advent of the knowledge of the particle nature of fluids, the equation of state for an ideal gas had been already known to be
 
 \[ pV = nRT = NkT \tag{1}, \]

where $R = 8.314 \: Joules \: per \:(mole - \degree K )$ is the Universal Gas Constant, $k = 1.381 \:\times 10^{-23} \: Joules \:  per \: \degree K$  is Boltzmann\textquotesingle s constant and $n$ is the number of moles of the gas, $n = {Nk \over R}$ , so, from \eqref{eq:ktpressure} we have

\[ pV = {2 \over 3} N \Big[ {1 \over 2} m {\overline {({\mathbf v})^2}} \Big] = NkT = nRT \tag{3} \label{eq:ktidealgas} \] 

Thus , from \eqref{eq:eqst} and \eqref{eq:ktpressure}, can be written, for example,

\[ T = {pV \over Nk} = {2 \over 3} {1 \over k}  \Big[{1 \over 2}m{\overline {({\mathbf v})^2}}\Big] \tag{4} \label{eq:kttemp} \]

where 

\[ k = {nR \over N} = {R \over \aleph} , \]

and 

\[ \aleph = 6.022 \times 10^{23} \]

is Avogadro\textquotesingle s number, the number of molecules in a mole.

The bracketed expression in (4) is interesting because it is just the average kinetic energy per molecule, again assuming that the gas is in equilibrium. 

The pressure and temperature expressed in terms of  the particle nature of the gas provides the link to the laws thermodynamics.
 
\begin{center}
**********************
\end{center}

As will be seen below, energy can be added to (or extracted from, in the case of refrigeration) the system through the manipulation of its temperature, pressure or volume. If the volume increases, work is done by the gas.  Work is done on the gas when the volume decreases. Pressure can only be manipulated by increasing the temperature or doing work on the gas. This is a verbal statement of the first law of Thermodynamics:

\[ \Delta U = Q - W \tag{5} \label{eq:firstlaw} \]. 

where $\Delta U$ is the total change in the system\textquotesingle s energy, $Q$ is energy added and $W$ is the work done $by$ the gas. Of course either $Q, \: W$ or both  and  can be negative. $Q$ is negative in a refrigeration process and $W$ is negative in a process of compression. Each of these can be controlled from the outside.  Only $p$ is an effect.  It can only be changed by manipulating $Q$ and $W$.

This development has thus shown that in suitable units and in $equilibrium$, temperature is just the average kinetic energy of each of the individual molecules, which is proportional to the average of the squared velocities of the molecules.

The kinetic energy per unit volume of the molecules is

\[ {U_{Total} \over V} = {N \over V}  \big({1 \over 2} m {\overline {({\mathbf v})^2}}\big) = {1 \over 2} \rho {\overline {({\mathbf v})^2}} = {3 \over 2} \rho {kT \over m} \]

and

\[ {\overline {({\mathbf v})^2}} = 3{kT \over m} \]

so

\[ {K.E.}_{avg. \ per \ molecule} = {3 \over 2} kT \]

\[ U_{Total} = {3 \over 2} NkT = {3 \over 2}nRT \tag{6} \label{eq:utotal} \]

or by using \eqref{eq:ktidealgas}, 

\[ U_{Total} = {3 \over 2}pV. \]

\subsection{Specific Heat Capacity}

\[ C = {Q \over {\Delta T}} = {\Delta U + W \over \Delta T}, \]

where $C$ is called the heat capacity, $Q$ is the heat added and $W$ is the work done $by$ the gas.

The molar specific heat capacity of an ideal gas is, in metric units, the number of joules of energy per mole that will be added when the gas temperature is increased by one degree Kelvin.  

There are two specific heats: the specific heat at constant volume, $C_V$, and the specific heat at constant pressure, $C_p$. 

According to \eqref{eq:utotal} molar the specific heat of a monatomic\footnote{More complex molecules can store energy internally in the form of oscillations or rotation and thus the expressions for their specific heat capacities are more complicated.} ideal gas at constant volume, i.e. when no work is done, then, should be just 

\[ C_V = {3 \over 2}R. \]

For such a gas the molecules have no internal degrees of freedom to contribute to the specific heat.  This agrees well with experiment.

If when energy  is added and the pressure is kept constant rather than the volume then work will be done by the gas in the amount of 

\[ W = p \times \Delta V. \]

From equation \eqref{eq:ktidealgas} there results,

\[ p \times \Delta V + V \times \Delta p = nR \times \Delta T \tag{7} \label{eq:derideal} \]

If $p$ is held constant then the second term on the left above is zero and all the energy added through $dT$ results in work and,

\[ { W \over \Delta T}= nR. \]

If, on the other hand, the volume is held constant then the pressure must increase and:

\[ {\Delta p \over \Delta T} = {nR \over V}. \]

The molar heat capacity of a gas is defined as,

\[ C = {{\Delta Q}\over {n \Delta T}} = {{\Delta U + W} \over {n\Delta T}}, \]

using the first law in the form \eqref{eq:firstlaw} above, i.e. $W$ is the work done by the gas.

There are two heat capacities, one for each of the terms in the left in equation \eqref{eq:derideal}, i.e. for the gas at constant volume and for the gas under constant pressure.  The specific heat capacity at constant volume is

\[ C_V = {{\Delta Q} \over {n\Delta T}} = {{\Delta U } + W \over {n\Delta T}}. \]

But since the volume is constant, $W = 0$ so

\[ C_V = {\Delta U \over n \Delta T} = {{ 3 \over 2} nR \Delta T \over n \Delta T} = {3 \over 2} R, \]

where, again, the volume is kept constant and therefore no work is done either by or on the gas and the entire change in energy is due to heating or cooling.

On the other hand, at constant pressure, the system is heated as before, but now it is also allowed to expand at constant pressure.  The heat capacity at constant pressure, then, has two terms: one for the temperature increase and one for the work done by the gas.  So the specific heat capacity at constant pressure is

\[ C_p = {\Delta U + W \over n \Delta T} = {{3 \over 2} nR \Delta T + p \Delta V \over n \Delta T} \]

but, at constant pressure the volume is allowed to change and

\[ p \Delta V = nR \Delta T, \]

so 

\[ C_p = {(C_V + R) n\Delta T \over n\Delta T} = {5 \over 2}R. \]

This derivation assumes that the gas is in equilibrium.  However in practice, the operation of a heat engine for example, pressure, volume and temperature will be functions of position and time. In this case the gas is always in turbulent flow. 

The control of these variables is effected only by manipulating the boundary conditions. One has no ability to change the laws under which a thermodynamic process evolves. In a heat engine the shape of the piston crown, the location and angles of the intake and exhaust valves, the shape of the combustion chamber, the stroke of the piston and the choice of fuel according to its chemical properties are variables that an engineer can control.   These boundary conditions are very important in the design of high-performance gasoline engines for example.  The control of flow patterns in the combustion chambers and even in the exhaust systems of Formula I automobile engines is important in the maximization of power output.  

In the design of heating and cooling systems also, one must take into account the convection currents produced by the systems.  These are controlled by designing the shapes and placement of air ducts and fans.   The Grands Orgues, big classical organs, utilize the laws of acoustics by the design of the various pipes. Diameters, pipe mouths, the alloys or the wood selected for the pipes and other parameters are carefully chosen. Other factors that are exploited are the variables of the space around the organ. All of these behaviors depend on acoustic behavior outside the realm of equilibrium.

\section{The Influence of the Philosophy of Science on Research }\label{woodsapp}

\begin{center}
L.C. Woods \\
(Balliol College, University of Oxford) \\ 
Lecture delivered at the Auckland Institute of Technology on Friday, 7th August, 1998. 
\end{center}

\subsection{Introduction}

When I was young, I read an article that claimed that the purpose of scientific theory was to $describe$ phenomena rather than to $explain$ it. This surprised and disappointed me. Apparently we could achieve no more than an empirical account of the real world, and could not expect to understand it. This conservative philosophy is known as $instrumentalism$, because it maintains that a theory is no more than an `instrument` for making predictions. The opposing view is that theories relate to underlying $mechanisms$ and that these are responsible for the observed phenomena. To know the mechanism is to understand the phenomenon. However this $realist$ philosophy usually depends on some $metaphysical$ elements, introduced to enlarge the fabric of the hidden world and thus to aid explanation. And it was the liberal introduction of such unobservable elements that added force to the instrumentalist position, an extreme form of which is known as $positivism$. This holds that all statements other than those describing or predicting observations are meaningless (including this statement?). Knowledge is only what can be verified directly. $Logical positivism$ augments positivism by admitting mathematical entities logically connected to observations, even if they are not directly measurable.
 
An alternative title for this talk would be `The legacy of Logical Positivism`. This philosophy was greatly boosted in the first half of this century by the difficulty of giving the formalism of quantum mechanics an objective and realist interpretation. Hidden variables of one sort or another, even including parallel universes, have been advanced as possible solutions. But the experts have not been convinced. That debate continues and I expect that it will ultimately be resolved in favour of a $realistic explanation$ rather than an $instrumentalist description$. In any case this unresolved difficulty with realism at the smallest scales does not justify our rejecting it at the deterministic mesoscopic and macroscopic scales. What I shall illustrate is the way in which some scientists` preference for mathematical description over physical explanation has led to important differences in the way they have pursued their research. Fields that are particularly vulnerable to the legacy of positivism are thermodynamics and plasma physics. Very often the researchers are unaware that they are labouring under the influence of a largely discredited philosophy.

\subsection{Some Philosophical Background}

By a `mechanism` I mean the representation of a real process in terms involving familiar physical actions, e.g. we might say that the thermoelectric transport of heat is due to the fact that higher energy electrons have a smaller probability of colliding with ions than those of lower energy, or that a solar prominence is supported high in the corona by magnetic forces. Once we have the mechanism identified, fitting it out with suitable mathematics is often the easier task. In specifying a mechanism, the first objective is to try to identify those features essential to the phenomenon under consideration. Elaboration of the model can follow when the minimalist position has proved itself. But it may be necessary to add some unobservable structure to the mechanism from the outset, e.g. the magnetic field lying out-of-sight well below the photosphere when modelling a sunspot structure. 

Ockham was a 14th-century, Scholastic philosopher, who attacked the supremacy of Papal power. His 
`razor` was the statement that entities are not to be multiplied beyond necessity. In a similar vein, Ernst Mach (1838 - 1916) stated that `it is the aim of science to present the facts of nature in the simplest and most economic conceptual formulations`. This was a reaction against the metaphysical extravagances of the 17th and 18th centuries, during which, inter alia, various fluids were adduced to `explain` physical phenomena. A classical case was the chemists` phlogiston that, having negative weight, supposedly explained the increase in mass due to burning -- the heat drove off the phlogiston. Oxygen was yet to be discovered. Heat had the properties of an indestructible fluid called `caloric`; electricity was said to be composed of two fluids, with no more evidence than this seemed to provide an explanation of some observations. But nowadays we do talk of electron and ion fluids. Also Carnot managed to establish the principle that later evolved into the second law of thermodynamics, by employing the caloric concept. (It was the first law that later destroyed the conservation of caloric.) So some metaphysical inventions prove to be closer to the truth than at first imagined. Such elements evolve from being metaphysical to eventually being considered to be `real`. 

Atoms were ruled out by Mach and other anti-atomists of his day. They could not be observed, so were not real. They could be admitted only as a device to give economy of thought. Realists are much bolder, willing to introduce unobserved elements and to take them as being real, in order to provide `explanations.` A classic example is William Harvey\textquotesingle s (1578-1657) explanation of the circulation of the blood and the function of the heart as a pump. Although he had no microscope to see the capillary vessels connecting the arterial and venous systems, he maintained from evidence implying a circulation that they must exist. Pauli\textquotesingle s (1930) invention of the neutrino to ensure the conservation of energy and momentum during beta decay is another good example of a bold metaphysical creation. It was not until powerful nuclear reactors were available that the existence of neutrinos could be confirmed. 

Analogy is a powerful, heuristic means of arriving at a possible description of a new phenomenon. It may be a mathematical likeness only, as with the fact that the temperature in steady-state heat flux and the gravitation potential both satisfy Laplace\textquotesingle s equation, or it can be a deeper physical analogy, such as that between the transport of heat by colliding particles and its transport by photons within the Sun. For example Einstein\textquotesingle s interpretation of Brownian motion as being due to the uneven bombardment of microscopic particles by molecules may have taken root in his mind from an obvious macroscopic analogy, and its success led quickly to the full acceptance of the reality of molecules. Metaphysical inventions have a central role in science, provided one always remembers that they are on `trial` until the indirect evidence is so strong that they can be considered to be `real`. 

Maxwell was pre-eminent in his use of analogy. He deployed it in two famous examples. First in his kinetic theory of gases he used a `billiard ball` model to describe the trajectories of the molecules. This was excellent for monatomic molecules, but partially failed with diatomic molecules for reasons that are now obvious to us. His other great analogy was to represent magnetic fields as vortex filaments in a `fluid`, and to separate them by particles each revolving on its own axis in the opposite direction from that of the vortices. These `idle-wheels` (later identified as `electrons`) were to allow the free rotation of the vortex filaments. To cap it all, he gave the vortices elastic properties to represent the displacement current! In this manner he arrived at his set of equations for the electromagnetic field. He found that the velocity with which disturbances propagated through his system of vortices and particles (70,843 leagues per sec.) agreed so closely with Fizeau\textquotesingle s value for the speed of light that he remarked ''we can scarcely avoid the inference that \textit{light consists in the transverse modulations of the same medium which is the cause of electric and magnetic phenomena.``} 

Then the mechanical description was abandoned, it being assumed that the equations $alone$ now represented the phenomenon. Maxwell\textquotesingle s analogy, however absurd it seems today, led him to the greatest discovery of the 19th century, namely that light was electromagnetic in nature. The French positivist, Pierre Duhem, observed sarcastically that Maxwell had cheated by falsifying one of the equations of elasticity in order to obtain a result that he already knew by other means. 

However if one asserts the whole theory is simply the equations, one is adopting a positivist point of view. It is Faraday\textquotesingle s great metaphysical construction -- the notion of an electromagnetic field permeating space -- that allows a return to a physical description of Maxwell\textquotesingle s equations. This continuum picture is very helpful in trying to understand the interaction of fields and particles. I shall use `mechanism` in this extended sense in what follows. 

Since logical positivists eschew physical mechanisms, they are attracted to mathematical treatments, especially when an axiomatic basis can be adopted or devised to give the approach the gloss of pure mathematics. The belief that, excepting blunders, mathematical proofs are absolutely certain and therefore superior to physical arguments, should have been dealt the \textit{coup  de gr{\^ a}ce} by G{\" o}del\textquotesingle s incompleteness theorem. This states that if a set of rules of inference in a branch of mathematics is consistent, then within that branch there must exist valid methods of proof that these rules fail to identify as valid. 

Of course the equations of mathematical physics $follow$ from the assumed mechanisms, but sometimes this dependence is inverted or forgotten and the equations begin to assume an independent significance well beyond their original range of validity. This does not seem to concern cosmologists, whose big bang, in which the universe is created from nothing via a quantum mechanical tunneling process, is a wild, but apparently successful, extrapolation of known physical laws. 

The importance of imagining phenomena in terms of mechanisms rather than the equations employed to represent them, is that mechanisms are often much more suggestive of modifications and extensions to more accurate models of the processes, whereas equations, especially if they are complicated as in the integro-differential equation of kinetic theory, or as in the full set of MHD equations, are less helpful. Equations sometimes have many terms, each of which usually represents a distinct physical process. It is important to try to relate the terms $individually$ to features of the physical model and not simply to lump them together. 

Phenomena in fields like biology, where the mechanisms are obscure or unknown, and which rely on statistical data to suggest causal connections, remain fertile for the positivists. To be true to their philosophy, they would be content to rest the case for the dangers of cigarette smoking on the correlation discovered between smoking and various types of illness. But the merchants of death are realists. The cigarette manufacturers insisted that, in the absence of proven biological $mechanisms$ relating disease to smoking, their product was innocent. It is true that mere correlation proves nothing, a classical case being the noted correlation between the incidence of prostitution in London just after the second World War and the salary of Bishops. 

\subsection{How Positivists Confused the Basis of Plasma Physics} 

Whether or not the early plasma physicists knew any philosophy of science I cannot tell, but it would appear from their mistakes that they preferred formal mathematics to physical mechanisms. The most obvious example is that plasma pressure was defined as momentum flux, which is correct only if molecular collisions are sufficiently numerous. The classical case of wall $pressure$ being due to its bombardment by molecules should have made the role of collisions obvious enough, but the $momentum \: flux$ definition, in which the collisions are only $implicit$, confused quite experienced scientists into believing that there could be pressure gradients even in a $collisionless$ plasma. One wonders by what mechanism can purely $ballistic$ particles transmit a pressure force. 

I was once challenged at a seminar I was giving at the UK Culham Laboratory for Fusion Research. A scientist claimed that `Collisions are $not$ essential for there to be a pressure gradient in a plasma!` His argument was that one could have a gradient in the number density, $n$, of ionised particles, maintained by a strong magnetic field (which is correct) and if the medium were isothermal, from the law $p \propto nT$ relating the pressure to the temperature T, it follows that there would be a gradient in the pressure. It seemed plausible to the audience. They had become so familiar with the classical pressure/temperature law that they had forgotten that its derivation required collisions, i.e. it is not true that $p \propto nT$ in a collisionless plasma. One could take it to be a definition, but then it would have no physical content.

Although this mistake seems harmless enough, it was compounded into a more ridiculous and even dangerous notion for a plasma in a strong magnetic field. If the magnetic field strength $B$ say, has a gradient in a direction orthogonal to the field vector ${\mathbf B}$, as the charged particles gyrate about the field lines with a radius inversely proportional to $B$, the variation in the radius of gyration experienced by the particles causes them to drift in a direction orthogonal to both ${\mathbf B}$ and its gradient. (The motion has a similarity to that of a top on an inclined plane -- the gravitational force down the plane results in a motion of the top along the plane at right angles to this force.) This is known as `grad B drift` and depends on the assumption that the average time interval $\tau$ between successive particle collisions is much greater the gyration time, $\omega^{-1}$. The individual particles have an average velocity ${\mathbf u}_{mathbf B}$ across the field determined by the value of $grad \: B$.

 Now consider the whole collection of particles treated as being a $fluid$. The equation of fluid motion does not have a term involving $grad \: B$, but it does have a term proportional to the pressure gradient, which gives rise to a fluid velocity ${\mathbf v}$ across the field lines depending on the magnitude of $grad \: p$. Now the fun begins. The average drift velocity ${\mathbf u}_{mathbf B}$ it would seem, cannot possibly be the same as ${\mathbf v}$, since the former depends only on $grad \: B$ and the latter only on $grad \: p$. This means that, with appropriate choice of the two gradients, it is possible to send the ion mass qua $particles$ in the opposite direction to the ion mass qua $fluid$. But there can be only $one$ direction of mass motion. The amazing thing is that this reputed `paradox` is acknowledged and accepted in the literature. That there $must$ be an error is not even appreciated. The simple mistake is to assume that in the guiding centre description the particles are collisionless, but in the fluid description they respond to pressure forces, i.e. to the impact of other particles. 

Why do I think this confusion over pressure is serious? Well the extremely expensive and unsuccessful fusion energy project, in which very hot plasma was supposed to be confined by magnetic fields, was based on theory in which guiding centre motion plays a role. And the (incorrect) equations resulting from not understanding the nature of plasma pressure failed to reveal that there would be a disastrous loss of plasma across the tokamak fields. But the error goes deeper than that. The basic equation on which all of the kinetic theory of plasmas was developed is Boltzmann\textquotesingle s kinetic equation, which we shall next consider.   
 
 \subsection{Why Boltzmann\textquotesingle s Equation is Incorrect}
 
 In 1872 Boltzmann published the paper ''Further Studies on the Thermal Equilibrium of Gas Molecules`` that contains his famous integrodifferential equation for the evolution of the density of particles in phase space. At a meeting in Vienna to commemorate the centenary of this publication, G.E.Uhlenbeck stated: 
 
 \begin{quotation}
 
The Boltzmann equation has become such a generally accepted and central part of statistical mechanics, that it almost seems blasphemy to question its validity and to seek out its limitations. It is also almost a miracle how the equation has withstood all criticisms...

\end{quotation}

However, that there is an important limitation to the equation becomes evident when terms second-order in the ratio of the microscopic (molecular collisional) time-scale to the macroscopic time-scale, known as the Knudsen number, are examined. (A typical second-order term involves two gradients, e.g. the heat flux includes a term   ${\mathbf q}_2 =  -\alpha \nabla {\mathbf v} \dot \nabla T$, where $\alpha$ is a constant and ${\mathbf v}$ is the fluid velocity. The first-order theory yields the classical transport equations of Fourier, Ohm and Newton and is correct.) For example there are physically evident terms for the heat flux across magnetic fields that $cannot$ be derived from Boltzmann\textquotesingle s equation. On the other hand, his equation leads to second-order terms for heat transport in an $isothermal$ neutral gas, in which circumstances no such transport is possible. One such term has $\nabla p$ in place of $\nabla T$ in the (correct) second-order expression for ${\mathbf q}_2$ just quoted. The physical nature of energy transport in a non-conducting gas $must$ depend on there being a temperature gradient. 

The fault with the equation lies with the assumption that the collision rate between molecules is proportional to the product of the distribution functions of the colliding particles, regardless of anisotropies generated by the presence of pressure gradients and fluid shear. Boltzmann\textquotesingle s main purpose was to find a way of deducing the second law of thermodynamics from mechanics and also to improve on Maxwell\textquotesingle s derivation of the equilibrium distribution. In this he certainly succeeded. In fact only seven of the 96 pages of his paper deal with the calculation of the transport properties of gases. The tacit assumption that the equation was valid over a wide range of Knudsen numbers, made by Chapman, Enskog and many others since, is where the fault really lies. 

  In the formulation of the kinetic equation, the distinction between convection and diffusion is not correctly drawn. Diffusion is due to molecular agitation superimposed on a reference frame that not only has the speed of the fluid element, but which also $accelerates$ and $spins$ with it. This means that the pressure gradient must be assumed known from the outset, since the fluid element to which the frame is attached, is accelerated by this force. The spurious terms in the heat transport mentioned above, arise because of the neglect of these fluid accelerations. For example, scattering is taken without comment to be isotropic in a frame that has the velocity of a fluid element, but not its acceleration. Had the mechanism on which the equation was based, namely the collisional scattering of molecules in and out of a $fully$ convected element of phase-space, been clearly understood and kept in mind, the error would have been soon discovered and the original Boltzmann\textquotesingle s equation corrected for higher values of the Knudsen number. 

Great sums and scientific effort have been invested into finding computer solutions of Boltzmann\textquotesingle s equation in the regime of large Knudsen numbers. Such a waste and just because of the apparently unshakeable belief in equations rather than mechanisms. So far as research in fusion energy is concerned, the cost of the failed tokamak machine world-wide must exceed ten billion dollars. At least in part this waste can be attributed to the propensity of plasma physicists to adopt a positivistic view of their science. 

\subsection{Understanding Entropy}

My final example of the legacy of positivism comes from that will-o`-the-wisp known as $entropy$. What is curious about this property of macroscopic systems is that it is a purely $defined$ quantity, not relating to a physical property until precise details of the $state$ of the system have been specified. And the problem with `state` is that it is $observer-dependent$, i.e. it depends on what elements the observer wishes to include in his physical model of the system under consideration. The greater the detail, the smaller the resulting entropy of the system. In this case the `mechanism` is simply the chosen physical state. Unfortunately the positivist position seems to be that entropy is a property of systems, independent of the observer -- the mechanism via state is quite ignored. Examples of this are to be found in a subject optimistically termed `rational thermodynamics`, in which temperature and entropy are taken to be `primitive` quantities not requiring definition. 

This ignorance would be harmless enough, except that `the` entropy, or rather its rate of production, say, is made the basis for determining the form of constitutive relations, namely the laws relating fluxes and the thermodynamic forces driving them. The principle is that these fluxes and forces appear as quadratic products in $\sigma$, which it is assumed must always be positive. The second law of thermodynamics is claimed to support this view. Fourier\textquotesingle s law relating the heat flux $q$ to the temperature gradient $\nabla T$ , is a simple example. In the absence of other processes, the production rate is $\sigma = -{\mathbf q} \dot \nabla T > 0$, which implies that ${\mathbf q } = - \kappa \nabla T$, where $\kappa$ is a positive constant termed the `thermal conductivity`. The $physical$ mechanism that generates these fluxes and that is responsible for $generally$ being positive is lost in a pseudo-mathematical haze, with the hope no doubt that the pure mathematical appearance of the formalism will impress the followers into believing that the reasoning is unassailable. 

Of course this approach leads to gross errors, perhaps the simplest of which is a theorem due to Coleman that asserts the entropy to be independent of the gradients of temperature and fluid velocity. In a theory correct to second-order in the Knudsen number, this is readily shown to be wrong. Another evident error arises in the theory of heat flux across strong magnetic fields. There is an interesting and very important term ${\mathbf q}_{\mathbin{\char`\^}}$ that happens to be orthogonal to $both$ the temperature gradient and the magnetic field. Since ${\mathbf q_{\mathbin{\char`\^}}} \dot  \nabla T = 0$, this term does not generate entropy and does not appear in the expression for $\sigma$. It therefore does not exist for the rational thermodynamicist. In fact it is the second-order form of ${\mathbf q}_{\mathbin{\char`\^}}$ that is responsible for the failure of tokamaks to retain their energy for more than a few seconds at best, when minutes would be required for an economic fusion reactor. And the same transport equation plays a dominant role in coronal physics. 

The entropy production rate is assumed to be always positive. This is presented as an axiom, with the second law as justification. Unfortunately the second law gives no guarantee that $\sigma$ is always positive. In a strong magnetic field it is in fact the case that $\sigma$ may have either sign. The argument is as follows. We expand $\sigma$ as a power series in the Knudsen number, $\epsilon$ say, 

\[\sigma \approx \sigma_1 + \sigma_2 \hspace{3em} (\sigma_i = \mathcal{O}(\epsilon^i)) \]

where we have carried the expansion only to the second-order term. Of the two terms, only $\sigma_1$ is dissipative. It is therefore always positive. The second-order term has three gradients and is consequently reversible, i.e. it may have either sign. Moreover it can be much larger than $\sigma_1$, in which case the total $\sigma$ may be negative. This is not a failure of the second law, which relates only to dissipative terms. The problem is that the rational thermodynamicist has adopted an axiom, the physical content of which he does not understand. 

It can be shown that the stability of a continuum flow requires that 

\[ \sigma_1 \geq 0, \hspace{4em} \sigma_1 + \sigma_2 \geq 0, \]

the inequalities in which have different roles. The first determines that the thermal conductivity and the fluid viscosity be positive quantities, making it a $thermodynamic$ constraint; it is universal and independent of the actual flows obtained. The coefficients of the second-order terms, being determined by the same molecular behaviour as the first-order coefficients, are closely related to them, leaving no freedom for further thermodynamic constraints to be satisfied. 

On the other hand the stability constraint $((\sigma_1 + \sigma_2) \geq 0)$ can be satisfied only by restricting the class of fluid flows. It sets a limit on the gradients, or equivalently on the Knudsen number making it a macroscopic or $fluid$ constraint. Provided $\epsilon$ is sufficiently small -- certainly less than unity -- $\sigma_1$ will be larger than $\mid \sigma_2 \mid$ and the constraint will be satisfied. We might imagine some initial state, with steep gradients, for which this does not hold. The fluid flow may be momentarily unstable, but the resulting fluctuation will quickly restore a new equilibrium state, in which $\sigma_2$ is either positive, or smaller in magnitude than $\sigma_1$. 

Since the second law of thermodynamics is not normally associated with the question of the stability or not of the flow field, it is restricted to the thermodynamic constraint,$\sigma_1 \geq 0$, while the constraint on $(\sigma_1 + \sigma_2)$ is concerned solely with the $macroscopic$ stability of the flow field. Certainly with magnetoplasmas, even with convergent Knudsen number expansions, the $(\mathcal{O}(\epsilon^2))$ terms dominate the classical $(\mathcal{O}(\epsilon))$ terms by orders of magnitude and instabilities in which $(\sigma_1 + \sigma_2)$ changes sign periodically do occur. The adoption of the inequality ?$\sigma > 0$ as an axiom is a serious mistake. 

\subsection{Conclusions}

I have tried to show that the attitude of scientists to their research -- the way they go about it -- is greatly influenced by the beliefs they have adopted, consciously or otherwise, about the nature of the scientific enterprise. In this way, the philosophy of science $does$ play an important, but indirect role in research, a point of view that would have been quite obvious to most 19th Century scientists. The importance of philosophical concerns in cosmology and quantum physics, in which branches of science there is no shortage of metaphysical invention, is obvious enough, but that this extends into the classical realms of continuum physics is not sufficiently appreciated. 

University teaching is largely responsible for the inculcation of positivist attitudes in the typical university graduate. Mathematical approaches to physics and the mathematical sciences are easier to teach and easier to learn. Mechanisms are not ignored, but they are given less attention, especially where it counts for the average student, namely in the examination room. When I was an examiner for Oxford Finals in Mathematics, I always attempted to set some questions essay-type discussions of underlying physical principles. But these were seldom answered -- they were thought to be too difficult and in any case how can one get high marks for a mere essay!

\end{document}